\documentclass[a4paper]{article}
\usepackage[a4paper]{geometry}
\usepackage[authoryear,longnamesfirst]{natbib}
\usepackage{bm}
\usepackage{subcaption}
\usepackage{amsmath}
\usepackage{amsfonts}

\usepackage[colorinlistoftodos,textsize=tiny,obeyFinal]{todonotes}

\newcommand{\LI}{{\bf L}^{\rm{I}}}
\newcommand{\LII}{{\bf L}^{\rm{II}}}
\newcommand{\LIII}{{\bf L}^{\rm{III}}}
\newcommand{\RD}{\mathrm{RD}}
\newcommand{\env}{\mathrm{env}}
\newcommand{\num}{\mathrm{num}}
\newcommand{\MS}{\mathrm{MS}}

\usepackage{xcolor}
\usepackage[normalem]{ulem}

\title{On the speed of propagation in Turing patterns for reaction-diffusion systems}
\date{version: \today}
\author{
    V{\'a}clav Klika$^{1}$ \and  Eamonn A. Gaffney$^{2}$ \and Philip K. Maini$^{2}$}
\date{%
$^{1}$Dept of Mathematics, FNSPE, Czech Technical University in Prague, vaclav.klika@fjfi.cvut.cz,\\%
    $^{2}$Mathematical Institute, University of Oxford,\\ gaffney@maths.ox.ac.uk, maini@maths.ox.ac.uk\\[2ex]%
    \today
}

\begin{document}
\maketitle
\begin{abstract}
This study investigates transient wave dynamics in Turing pattern formation, focusing on waves emerging from localised disturbances. While the traditional focus of diffusion-driven instability has primarily centred on stationary solutions, considerable attention has also been directed towards understanding spatio-temporal behaviours, particularly the propagation of patterning from localised disturbances. We analyse these waves of patterning using both the well-established marginal stability criterion and weakly nonlinear analysis with envelope equations. Both methods provide estimates for the wave speed but the latter method, in addition, approximates the wave profile and amplitude. We then compare these two approaches analytically near a bifurcation point and reveal that the marginal stability criterion yields exactly the same estimate for the wave speed as the weakly nonlinear analysis. Furthermore, we evaluate these estimates against numerical results for Schnakenberg and CDIMA (chlorine dioxide-iodine-malonic acid) kinetics.  In particular, our study emphasises the importance of the characteristic speed of pattern propagation, determined by diffusion dynamics and a complex relation with the reaction kinetics in Turing systems. This speed serves as a vital parameter for comparison with experimental observations, akin to observed pattern length scales. Furthermore, more generally, our findings provide systematic methodologies for analysing transient wave properties in Turing systems, generating insight into the dynamic evolution of pattern formation.

\end{abstract}



\textbf{Keywords:}
  travelling waves; front propagation; pattern formation; marginal stability; envelope equation; pulled fronts

\section{Introduction}\label{sec:intro}
%





While patterns induced by self-organisation, for example by Turing's mechanism of diffusion-driven instability, are  traditionally considered in terms of stationary solutions \citep{murray2003mathematical},  there has also been extensive interest in spatio-temporal behaviours arising from such systems, for instance the propagation of patterning from a localised disturbance \citep{tarumi1989wavelength, myerscough1992analysis, liu2022control}. As a specific example, we consider the common example of Schnakenberg kinetics \citep{schnakenberg1979simple} on a one-dimensional spatial domain: 
\begin{equation} \label{eq:RD0}
	\frac{\partial \bm{u}}{\partial t} = {\bf D} \frac{\partial^2 \bm{u}}{\partial x^2} + {\bf R}(\bm{u}), ~~~~~~ \bm u = 
	\left( 
	\begin{tabular}{c}
		$u_1$ \\ $u_2$ 
		\end{tabular} \right) , \qquad {\bf R}(\bm u)  = 
	\left( 
\begin{tabular}{c}
$a-u_1+u_1^2u_2,$\\
$b-u_1^2 u_2 $ 
\end{tabular}  \right), \qquad {\bf D}  = 
	\left( 
\begin{tabular}{c c}
$D_{1}$ & $0$\\
$0$ & $D_{2}$ 
\end{tabular}  \right).
\end{equation}
Here $a,~b,~D_{1}$ and $D_{2}$ are positive parameters and the boundary conditions are zero flux. This system has one spatially homogeneous steady state
\begin{equation} \label{eq.Sch_HSS}
{\bf u}^*=\left(a+b, \frac{b}{(a+b)^2}\right).
\end{equation}

\begin{figure}[h]
\centering
\begin{subfigure}{.48\textwidth}
   \centering
   \includegraphics[width=.95\linewidth]{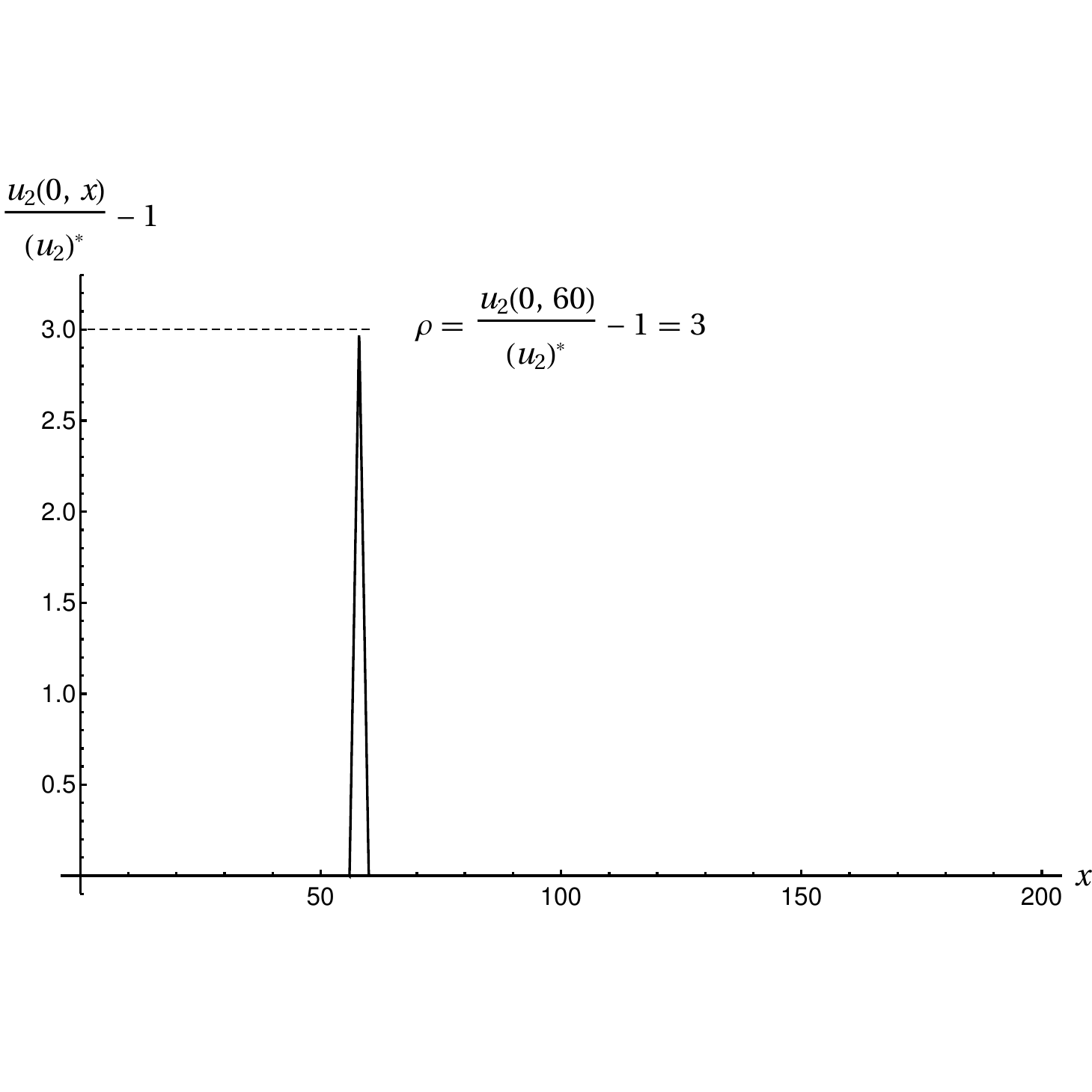}  
   \caption{A localised perturbation of the homogeneous steady state in the second component $u_2$, which is a piecewise linear function with a support of size $4$. It is localised at $x=60$ and given in multiples of the homogeneous steady state $u_2^*$ which we express using $\rho=u_2(t=0,x=60)/u_2^*-1$. In this case, we consider $\rho=3.0$.}
   \label{fig1:IC}
 \end{subfigure}
 \hspace{0.3cm}
\begin{subfigure}{.48\textwidth}
   \centering
   \includegraphics[width=\linewidth]{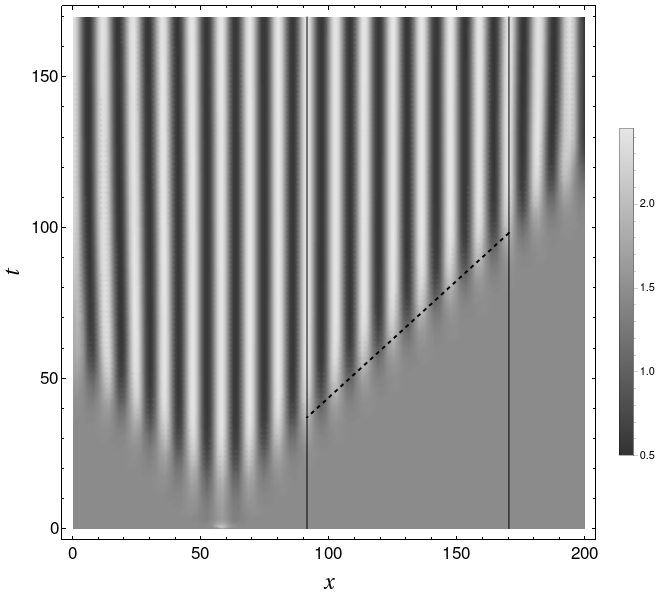}
   \caption{We initiate the simulation with a local perturbation in the second component as shown in the left panel. The density plot of the first component of the solution, $u_1$, shows the resulting transient wave of patterning spreading across the domain in both directions (left and right) with an approximately constant velocity, prior to the establishment of a stationary Turing pattern. After an initial transient period, we can indeed observe the asymptotic establishment of a constant travelling wave speed as indicated by the straight dashed line.}
   \label{fig1:Sch-PDEsolu}
\end{subfigure}
    \caption{\label{fig1}
 Illustration of pattern formation behind a wave travelling at a constant speed after a localised perturbation to the steady state. We consider Schnakenberg system, Eqn. \eqref{eq:RD0}, with zero-flux boundary conditions and parameter values $D_1 =1,~ D_2 =20$, $a=0.05,~b=1.4$ and a 1D domain with size $L=200$ (which we denote as Schnakenberg I). The profile of $u_{2}$ is out of phase with the profile of $u_{1}$.}
\end{figure}

For a localised perturbation about this steady state, a stationary Turing pattern  eventually forms via a transient pattern propagating across the domain  at an approximately constant speed 
after initiation, as highlighted in Figure \ref{fig1}. Here we focus on this transient wave and its properties, especially its asymptotic speed, amplitude and profile.
We also consider a modelling representation, albeit simplified, of the CDIMA (chlorine dioxide-iodine-malonic acid) reaction kinetics taken from \citet{konow2019turing}, which is based on the two-variable version of the kinetics \citep{lengyel1991modeling}, where the short range species $u_{1}$ inhibits the production of $u_{2}$, 
with the latter promoting the production of $u_{1}$. This constitutes a well known example of pure kinetics \cite{murray2003mathematical}, in distinct contrast to the cross kinetics of Schnakenberg, with the opposite interactions, whereby $u_{1}$ catalyses the production of $u_{2}$, which inhibits the production of $u_{1}$.




The speed of such a wave is amenable to the application of the marginal stability criterion of \citet{dee1983propagating,tarumi1989wavelength} and \citet{myerscough1992analysis}. Originally  formulated as a hypothesis \citep{dee1983propagating}, 
the application of the marginal stability criterion is subject to various  heuristics to motivate its use. Marginal stability has been motivated as corresponding to the transition between convective and absolute instabilities \citep{rovinsky1992chemical,tobias1998convective,sandstede2000absolute,sherratt2014mathematical,ponedel2017front} while it is argued that the application of marginal stability relies on the observation that there is a velocity of the moving reference frame at which the leading edge of a perturbation neither grows nor decays, as can be assessed by a Fourier representation of the solution in a fixed reference frame \citep{ben1985pattern}. 
The exact conditions under which this criterion can be applied are still an open question; furthermore, marginal stability provides no information on the amplitude or profile of the transient wave.   
Thus we  will also consider such a propagating front in terms of weakly non-linear analysis and envelope equations \citep{hoyle2006pattern}, which is justified by weakly non-linear analysis and multiple scale asymptotics,  and additionally can provide estimates for the wave shape and amplitude.
However, we acknowledge at the outset that both techniques are unlikely to be valid if a subcritical bifurcation is present; for instance marginal stability is considered to be limited to pulled fronts \citep{ponedel2017front,avery2023pushed}, while weakly non-linear analysis is not sufficient to resolve complexities such as the impact of multiple steady states influencing the dynamics.

With a focus on supercritical instabilities, the primary aim of this paper is to provide systematic 
methodologies  for determining the properties of   propagating patterning transients for Turing systems, such as those observed in Figure \ref{fig1}b. Noting that it is simpler to implement,  the marginal stability approach is presented first. 
Then we develop a weakly non-linear analysis to determine the envelope equation, which takes the form of a real Ginzburg-Landau equation (GLE) to determine estimates, not only for the wave speed, but also for the wave amplitude and profile, which we compare with numerical simulations.  In addition,  we also   compare wave speed predictions and analytical expressions from both methods, with the objective of ascertaining  whether  the simpler, more tractable, marginal stability method can provide similar accuracy to that of envelope methods for predicting the speed of patterning transients, and whether envelope methods can provide a quantitative justification for the use of marginal stability wave-speed estimates.

\section{Analytics - the foundations for determining the characteristic speed of patterning}

We consider a one-dimensional reaction-diffusion (RD) system of two species,
\begin{equation}
  \label{eq:RD}
  \frac{\partial \bm{u}}{\partial t} = {\bf D} \frac{\partial^2 \bm{u}}{\partial x^2} + {\bf R}(\bm{u},\alpha), 
\end{equation}
subject to homogeneous Neumann (zero flux) boundary conditions, where $\alpha$ denotes a bifurcation parameter such that $\alpha=0$ at a bifurcation point corresponding to the loss of stability of the homogeneous steady state (HSS), $\bm{u}^*(\alpha)$, i.e. $\bm{R}(\bm{u}^*(\alpha),\alpha)=0$. Without loss of generality, the choice of sign of $\alpha$ entails the HSS is stable for $\alpha<0$ and unstable for $\alpha>0$. We shall denote vectors as $\bm{u}$ and matrices as ${\bf A}$ throughout the text.

\subsection{Marginal stability}

The marginal stability approach \citep{dee1983propagating,ben1985pattern,vansaarloos1988front,tarumi1989wavelength,myerscough1992analysis,chomaz2000propagating} identifies the location in time and space where the instability arises in the form of a travelling wave due to a localised perturbation.
We define $g(k)=ikc+\sigma(k)$, where $k$ is the wavenumber and $\sigma(k)$ is the classical dispersion relation \citep{murray2003mathematical,krause2021modern,klika2017significance}. The marginal stability criterion then states that
\begin{equation*}
  0=\frac{d g}{d k},\quad \Re g = 0,
\end{equation*}
evaluated at the leading edge. However, the diverse arguments supporting the formulation of marginal stability conditions are all heuristic 
\citep{dee1983propagating,ben1985pattern,vansaarloos1988front,tarumi1989wavelength,myerscough1992analysis,chomaz2000propagating,vansaarlos2003front} and it is hypothesised that there is a change in the separation of variables of the solution from travelling wave (TW) coordinates to pattern-generating envelope fronts with fixed periodicity in space \citep{dee1988bistable}.



For the complex derivative $\frac{d g}{d k}$ to exist, we assume $g$ to be analytic in $k$, so that the Cauchy-Riemann conditions hold and the marginal stability conditions can be rewritten as
\begin{align}
  &\frac{\partial \Re\sigma}{\partial k^R} = \frac{\partial \Im\sigma}{\partial k^I} =0,\nonumber\\
  &\frac{\partial \Im\sigma}{\partial k^R} = -  \frac{\partial \Re\sigma}{\partial k^I} = -c,\label{eq.MargInst}\\
  & -c k^I +\Re \sigma = 0,\nonumber
\end{align}
where $k=k^R+i k^I$, $k^R,~k^I\in\mathbb{R}$. Note that $c$ decouples from this set of equations ($c=\frac{1}{k^I} \Re\sigma$) and we are left with two equations for the two unknowns $k^R,~k^I$.

In practical terms, equations \eqref{eq.MargInst} are solved as follows. First, the dispersion relation $\sigma(k)$ (recall that both $\sigma,~k\in\mathbb{C}$) is obtained as a solution to the following quadratic equation with complex coefficients
\begin{equation*}
 0= \det(\sigma{\bf I}+k^2 {\bf D}-{\bf J}) = \sigma^2+\sigma\underbrace{(k^2 \mathrm{tr}{\bf D}-\mathrm{tr}{\bf J})}_{p+iq}+\underbrace{\det(k^2{\bf D}-{\bf J})}_{r+is},
\end{equation*}
where ${\bf J}$ denotes the linearised kinetics, that is $J_{ij}=R_{i,u_j}$ and we denote partial derivatives via the comma notation, so that $R_{i,u_j}=\frac{\partial R_i}{\partial u_j}\Big|_{\bm{u}^*}$. Consequently, the following set of two equations (quadratic in $k$), implicitly defining the real and imaginary parts of $\sigma(k)$, is obtained:
\begin{align}
  \left(\Re\sigma+\frac{p}{2}\right)^2 -\left(\Im\sigma+\frac{q}{2}\right)^2 + \frac{q^2}{4}-\frac{p^2}{4}+r&=0,\label{Resigma_implicit}\\
  2 \left(\Re\sigma+\frac{p}{2}\right)\left(\Im\sigma+\frac{q}{2}\right)+s-\frac{pq}{2}&=0.\label{Imsigma_implicit}
\end{align}
This pair of equations, together with the first two equations in \eqref{eq.MargInst}, where we rewrite all the powers of $k$ in terms of its real, $k^R$, and imaginary, $k^I$, parts constitute a set of four relations for four unknowns. Finally, the travelling wave speed is given by $c=\frac{1}{k^I}\Re \sigma$ while the wavenumber of the pattern can be estimated as $k_{\MS}=\Im g/c = k^R+\Im\sigma/c$. The latter follows from the observation that the imaginary part of the shifted dispersion relation to the moving frame of the travelling wave is the frequency of the pattern scaled by the wave velocity $c$.


\subsubsection{Marginal stability close to bifurcation point}

In order to have a better understanding of the marginal stability approach, we asymptotically solve the marginal stability conditions close to the bifurcation point. To this end, we rescale the bifurcation parameter as $\alpha=\epsilon^2$, expand the linearised kinetics as ${\bf J}={\bf J}_0+\epsilon^2 {\bf J}_1+\epsilon^4{\bf J}_2$, where, for example, $(J_1)_{ij}=\frac{\partial^2 R_i}{\partial u_j\partial u_k}\Big|_{\bm{u}^*} \frac{\partial (u^*)_k}{\partial \alpha}\Big|_{\alpha=0}$. 

Standard linear analysis reveals that the bifurcation point is characterised by the condition
\begin{equation}
  \label{eq:bifPointCond}
(D_2 (J_0)_{11}+D_1 (J_0)_{22})^2-4 D_1 D_2 \det {\bf J}_0=0  
\end{equation}
and the critical wavenumber $k_c$ is given by the condition for a repeated root for $k^2$  \citep{murray2003mathematical}
\begin{equation}
  \label{eq:kc}
  k_c^2=\frac{D_2 (J_0)_{11}+D_1 (J_0)_{22}}{2 D_1 D_2}.
\end{equation}

As is usually the case with asymptotic solutions, the appropriate scaling can only be determined by calculation and the existence of a dominant balance in all considered orders. When considering regular asymptotic expansions in the small parameter $\epsilon$, this leads to (by trial and error) the choice of
\begin{equation}
  \label{eq:MSscaling}
  k^R=k^R_0+\epsilon^2 \kappa_R,\quad k^I=k^I_0+\epsilon \kappa_I,
\end{equation}
where $\kappa_R,~\kappa_I$ play the role of the unknown variables while we denote $\Re\sigma=\Re\sigma_0+\epsilon^2 \Re\sigma_1+\mathcal{O}(\epsilon^4)$ and $\Im\sigma=\Im\sigma_0+\epsilon^2 \Im\sigma_1+\mathcal{O}(\epsilon^4)$.

The nonlinear leading order problem is satisfied exactly at the bifurcation point, i.e. when
\begin{equation*}
  k^I_0=0,~k^R_0= k_c,~\Re\sigma_0=0,~\Im\sigma_0=0,
\end{equation*}
and, as a result, the leading order TW speed is zero. 

The first subleading problem is solved sequentially as follows. First, the implicit equations for the real and imaginary parts of $\sigma$, eqns. \eqref{Resigma_implicit} and \eqref{Imsigma_implicit}, yield
\begin{multline*}
  \Re\sigma_1=\frac{1}{2 (J_0)_{21} (D_1-D_2) (D_1 (J_0)_{22}-D_2 (J_0)_{11})}\\ \Big[D_1^2 (J_0)_{22} (2 (J_0)_{21} (J_1)_{22}-(J_0)_{22} (J_1)_{21})+D_2^2 (J_0)_{11} \left(2(J_0)_{21} \left(4 D_1 \kappa_I^2+(J_1)_{11}\right)-(J_0)_{11} (J_1)_{21} \right)\\
 +2 D_1 D_2 \left(2 (J_0)_{21}^2 (J_1)_{12}+(J_0)_{11} (J_0)_{22}(J_1)_{21} -(J_0)_{21} \left((J_0)_{11} (J_1)_{22}+(J_0)_{22} (J_1)_{11} -4 D_1 (J_0)_{22} \kappa_I^2\right)\right)\Big],
\end{multline*}
and $\Im\sigma_1=0$.

From the imaginary part of the first condition for marginal stability, $\frac{\partial \Im\sigma}{\partial k^R}=-c$, we obtain the following relation for speed $c$ (with the expected scaling with the distance from the bifurcation point):
\begin{equation*}
 c=\epsilon \frac{8  D_1 D_2 (D_1 (J_0)_{22}+D_2 (J_0)_{11})}{(D_1-D_2) (D_1 (J_0)_{22}-D_2 (J_0)_{11})}\kappa_I.
\end{equation*}
Further, the second condition, $\Re g=0$, gives the relation for $\kappa_I$:
\begin{multline*}
  \kappa_I^2 = \frac{1}{8 (J_0)_{21}D_1 D_2 (D_1 (J_0)_{22}+D_2 (J_0)_{11} )}  \Big[D_1^2 (J_0)_{22} (2 (J_0)_{21} (J_1)_{22}-(J_0)_{22} (J_1)_{21} )\\+2 D_1 D_2  \left( (J_0)_{11}(J_0)_{22} (J_1)_{21}+2 (J_0)_{21}^2 (J_1)_{12}-(J_0)_{21} (J_0)_{22} (J_1)_{11}-(J_0)_{11} (J_0)_{21} (J_1)_{22}\right)\\+D_2^2 (J_0)_{11} (2 (J_0)_{21}(J_1)_{11}-(J_0)_{11}(J_1)_{21})\Big].
\end{multline*}

Therefore, the marginal stability conditions close to the bifurcation point give the following estimate of the TW speed (in $t,~x$ dimensional coordinates)
\begin{multline}\label{eq.cMSBifP}
  c_{\rm MSasympt}=\pm \frac{\epsilon}{\sqrt{(J_0)_{21}} (D_1-D_2) (D_1 (J_0)_{22}-D_2 (J_0)_{11})} \Big[2 \sqrt{-2D_1D_2(D_1 (J_0)_{22}+D_2 (J_0)_{11})} \\\times \Big(D_1^2 (J_0)_{22} ((J_0)_{22}(J_1)_{21} -2 (J_0)_{21} (J_1)_{22})+D_2^2 (J_0)_{11}((J_0)_{11} (J_1)_{21}-2 (J_0)_{21} (J_1)_{11})\\+2 D_1 D_2 \left((J_0)_{11} (J_0)_{21} (J_1)_{22}-(J_0)_{11} (J_0)_{22} (J_1)_{21} -2 (J_0)_{21}^2 (J_1)_{12}+(J_0)_{21} (J_0)_{22} (J_1)_{11}\right)
    \Big)^{1/2} \Big].
  \end{multline}

To obtain the correction to the wavenumber and to check that we indeed have a plausible asymptotic solution with the assumed scaling, we calculate $\kappa_R$ from the real part of the first condition in \eqref{eq.MargInst} for marginal stability
{\small
  \begin{multline*}
    \kappa_R=\frac{\sqrt{2 (J_0)_{11}D_2+2 (J_0)_{22}D_1} }{8 \sqrt{D_1D_2}(J_0)_{21} (D_1-D_2) (D_1 (J_0)_{22}-D_2 (J_0)_{11}) (D_1 (J_0)_{22}+D_2 (J_0)_{11})}\Big[D_1^3 (J_0)_{22} \left((J_0)_{22} (J_1)_{21} -12 D_2 (J_0)_{21} \kappa_I^2\right)\\-D_1^2 D_2 \left(4 (J_0)_{21} \left(9 D_2 (J_0)_{11} \kappa_I^2+(J_0)_{22} \left(9 D_2 \kappa_I^2-(J_1)_{11}+(J_1)_{22}\right)\right)+(J_0)_{22} (J_1)_{21} (2 (J_0)_{11}-(J_0)_{22})+4 (J_0)_{21}^2 (J_1)_{12}\right)\\+D_1 D_2^2 \left(-2 (J_0)_{11} \left(2 (J_0)_{21} \left(3 D_2 \kappa_I^2+(J_1)_{11}-(J_1)_{22}\right)+(J_0)_{22} (J_1)_{21}\right)+(J_0)_{11}^2 (J_1)_{21}-4 (J_0)_{21}^2 (J_1)_{12}\right)+D_2^3 (J_0)_{11}^2 (J_1)_{21}\Big].
 \end{multline*}
}
Consequently, the wavenumber is estimated as 
{\small
\begin{multline*}
  \kappa^R+\Im \sigma/c= \sqrt{\frac{D_2 (J_0)_{11}+D_1 (J_0)_{22}}{2 D_1 D_2}}+\epsilon^2\sqrt{\frac{D_2 (J_0)_{11}+D_1 (J_0)_{22}}{D_1 D_2}}\\\times\frac{1}{8 \sqrt{2} (D_1-D_2)  (J_0)_{21} (-D_2  (J_0)_{11}+D_1  (J_0)_{22})( D_2  (J_0)_{11}+D_1  (J_0)_{22})^2}\Big[D_1^4 (J_0)_{22}^2 (5 (J_0)_{22}(J_1)_{21}-6(J_0)_{21}(J_1)_{22})\\+D_1^3D_2 (J_0)_{22} \left(2 (J_0)_{21} (-6 (J_0)_{11} (J_1)_{22}+7 (J_0)_{22} (J_1)_{11}-13 (J_0)_{22} (J_1)_{22})+(J_0)_{22} (J_1)_{21} ((J_0)_{11}+11 (J_0)_{22})-20(J_0)_{21}^2 (J_1)_{12}\right)\\+D_1^2 D_2^2 \Big((J_0)_{11}^2 (18 (J_0)_{21} (J_1)_{22} -17 (J_0)_{22} (J_1)_{21})+(J_0)_{11} \left(-44 (J_0)_{21}^2 (J_1)_{12}+12 (J_0)_{21}(J_0)_{22} ((J_1)_{11}+(J_1)_{22})-17 (J_0)_{22}^2 (J_1)_{21}\right)\\+2(J_0)_{21} (J_0)_{22}  (9 (J_0)_{22} (J_1)_{11}-22 (J_0)_{21} (J_1)_{12})\Big)\\+D_1 D_2^3 (J_0)_{11} \left(11 (J_0)_{11}^2 (J_1)_{21}+(J_0)_{11}(-26 (J_0)_{21} (J_1)_{11}+14 (J_0)_{21} (J_1)_{22}+(J_0)_{22} (J_1)_{21} )-4 (J_0)_{21} (5 (J_0)_{21} (J_1)_{12} +3 (J_0)_{22} (J_1)_{11})\right)\\+D_2^4 (J_0)_{11}^2 (5 (J_0)_{11} (J_1)_{21}-6 (J_0)_{21} (J_1)_{11})\Big],
\end{multline*}}
noting that the Cauchy-Riemann conditions are satisfied at both considered orders.

\subsubsection{Evaluation of marginal stability criterion performance}

We consider Schnakenberg and CDIMA  kinetics (as two representatives of kinetics which are in phase, CDIMA, and out of phase, Schnakenberg) to compare the marginal stability results with numerical solutions both near and far from the bifurcation point. To this end, we consider a primed Turing system, that is, we choose parameters from the Turing space (diffusion-driven instability region).

Both models involve two morphogens. The Schnakenberg kinetics are specified above in Eq. \eqref{eq:RD0} 
while the CDIMA kinetics are
\begin{subequations} \label{eq.CDIMA}
\begin{align}
  R_1(u_1,u_2)&=a-u_1-4 \frac{u_1 u_2}{1+u_1^2},\\
  R_2(u_1,u_2)&=\mu b \left(u_1-\frac{u_1 u_2}{1+u_1^2}\right),
\end{align}
\end{subequations}
where $a,~b,~\mu$ are (positive) model parameters.

For the case of zero flux boundary conditions, the spatially homogeneous stationary solution of the Schnakenberg model has been given in Eq. \eqref{eq.Sch_HSS} while, for the CDIMA kinetics, we have
\begin{equation*}
{\bf u}^*= \left(\frac{a}{5}, 1 + \frac{a^2}{25}\right).
\end{equation*}





\begin{figure}
\centering
    \includegraphics[width=.5\linewidth]{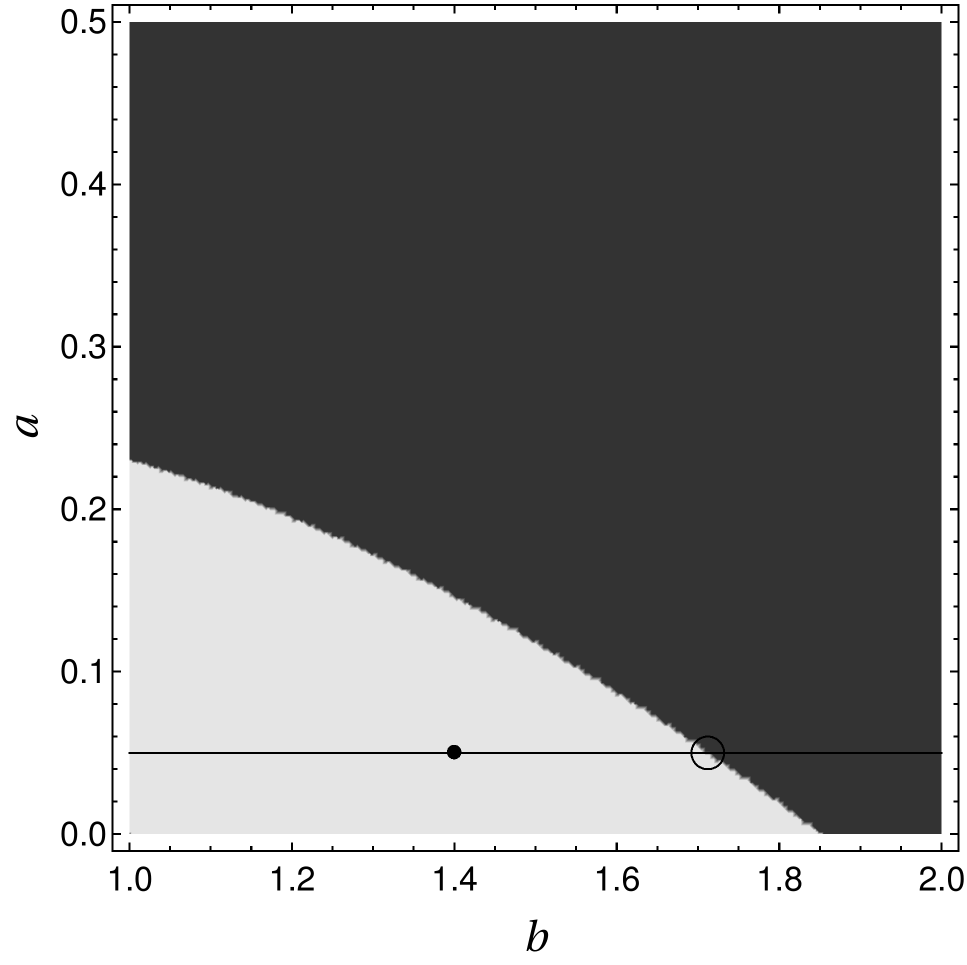}
    \caption{\label{fig2:Sch-Tspace} The Turing space, the set of points which allow Turing instability, for Schnakenberg kinetics with parameters $a,~b$, is highlighted in grey, where we fix the parameter values to $D_1 =1,~ D_2 =20$ and solve on a 1D domain with size $L=200$. The full dot represents the chosen point for numerical solution of the full problem (Schnakenberg I). The open circle denotes the closest bifurcation point $b_c=1.712$ (corresponding to $\alpha=0$) for the chosen bifurcation parameter $\alpha=b_c-b$.}
\end{figure}

We now consider the following set of parameters in the Schnakenberg kinetics above: $D_1 =1,~ D_2 =20$, $a=0.05,~b=1.4$ and a 1D domain with size $L=200$ (Schnakenberg I). This set of parameters is within the Turing space as indicated in Figure \ref{fig2:Sch-Tspace}, where we plot the Turing space in the $a,~b$ parameter space for $D_1$, $D_2$ fixed at $1, 20$ respectively. We also highlight the localisation of the nearest bifurcation point $b_c=1.712$ for the chosen bifurcation parameter $\alpha=b_c-b$, which leads to $\epsilon=\sqrt{\alpha}=0.559$.

\hspace{0.3cm}
\begin{figure}
    \centering
    \includegraphics[width=.5\linewidth]{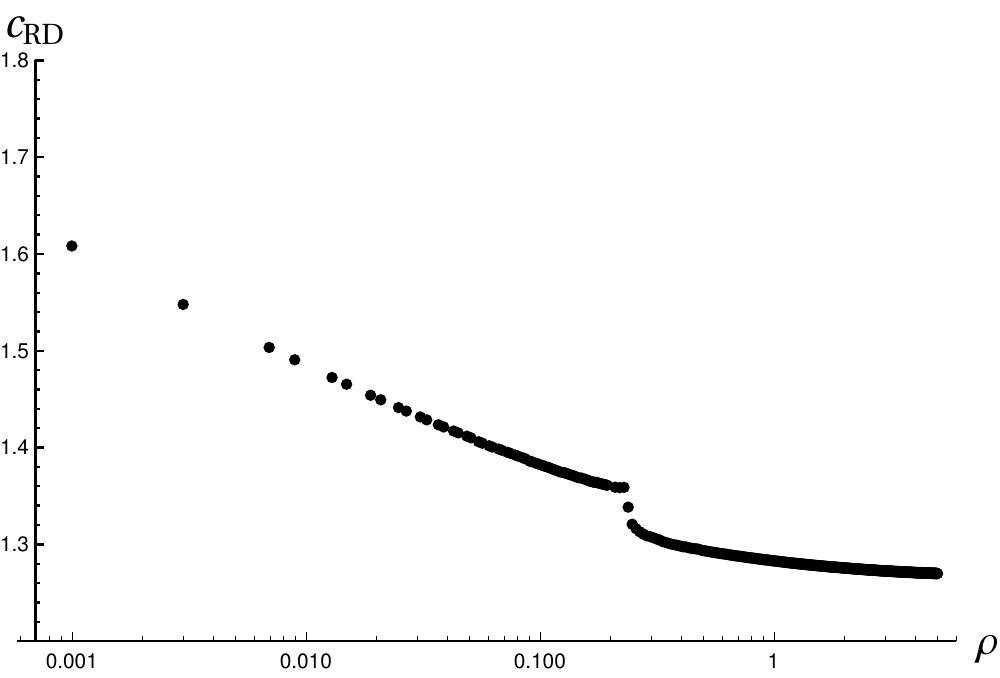}
    \caption{Investigation of the role of the magnitude of the local perturbation on the numerically observed travelling wave velocity. 
      The horizontal axis (in log scale) expresses $\rho>0$, the magnitude of the perturbation in the second component as described in Fig. \ref{fig1:IC}, while on the vertical axis we plot the numerically calculated $c_\RD$ using the same methodology as described in the text for the Schnakenberg I system. To this end, we use the two black vertical lines in Fig. \ref{fig1:Sch-PDEsolu} highlighting the locations of two cross-sections for the assessment of the pattern amplitude $A_\RD$, wavenumber $k_\RD$ and speed of the travelling wave $c_\RD$. Note that the intermediate transition around $\rho=0.2$ is due to finite size effects, when slight variations in $\rho$ cause an abrupt insertion of an additional half-wave of a pattern into the domain, that is, a slight variation of the pattern wave number $k_{\RD}$ occurs. }
    \label{fig3:Sch-CvsIC}
\end{figure}


We use Wolfram Mathematica 12.0 to solve the full PDE system with zero flux boundary conditions and a local perturbation of the homogeneous steady state ${\bf u}^*$ in the second component $u_2$ using the method of lines and 10,000 points for spatial discretisation. Pattern formation is initiated from a local perturbation well inside the domain, at $x=60$, and travels in both directions at an apparently fixed speed, see Figure \ref{fig1:Sch-PDEsolu}, in agreement with the previous observations in the literature \citep{liu2022control}. 

The choice of  magnitude for the local perturbation, as characterised by $\rho$ (see Fig \ref{fig1:IC}), impacts the duration of the transient time before the travelling wave velocity asymptotes to its final speed, as may be observed in Fig  \ref{fig3:Sch-CvsIC} for the Schnakenberg system in that the deviation between the travelling wave speed and its large $\rho$ asymptote in the plot is due to the duration of the transient. Since we solve the PDE numerically on a bounded domain and for different kinetics and various parameter values, the transient time will inevitably vary.
Thus, 
we opt for an initial condition of significant magnitude, as demonstrated in Fig. \ref{fig3:Sch-CvsIC}, such that the front velocity  has adequately  asymptoted within the considered bounded domain. In accordance with this objective, henceforth we select an initial perturbation for all numerical computations to be three times the homogeneous stationary solution in the second variable, that is $\rho=3$, with the perturbation of the first component remaining at zero. Then the numerically determined velocity can provide a reasonable estimation of the asymptotic rate of pattern propagating in the environment. Such an estimation can then be used for comparison.



From the numerical solution determined in this manner, we may extract all the observed characteristics from the (numerical) localisation of the highlighted maxima of the pattern, as in Figure \ref{fig1:Sch-PDEsolu}, with:
i) the amplitude $A_\RD=0.982$; ii) the pattern wavenumber $k_\RD=0.571$, denoting $2\pi$ divided by the distance between the neighbouring maxima; and iii) the speed of the travelling wave $c_\RD=1.288$ from numerically calculating  the times when the travelling pattern has reached half of the final amplitude at the two highlighted locations.
We now compare the results of the numerical solutions to those of the marginal stability analysis. To this end, we denote the predictions of the wavenumber and travelling wave speed from marginal stability by $k_\MS$ and $c_\MS$, respectively. Numerical solution of Eqns.~\eqref{eq.MargInst} for the above case gives $k_\MS=0.555$ and $c_\MS=1.375$ for Schnakenberg I, for example.

We repeat this process for other choices of parameter values and we also consider the CDIMA kinetics. The results are summarised in Table \ref{tab1MS}, where absolute and relative errors with respect to the numerical solution of both wave number and front speed are shown (more details for the CDIMA calculations are given in Appendix \ref{App.OtherCases}). Note that the longest transient time, and hence the greatest numerical error in calculated properties $k_\RD$ and $c_\RD$, can be expected near the bifurcation point, when the perturbation growth rate is very slow and hence the requirement for a sufficiently large region for determining the velocity of front propagation is greatest. 
Hence, the larger relative error near the bifurcation point is not necessarily a sign of a poorer performance of the marginal stability criterion. 

\begin{table}[h!]
  \begin{center}
  \begin{tabular}{ | l | c | | c | c | c | c | c | }
    \hline
    &parameters & $\epsilon=\sqrt{b_c-b}$ & $c_\MS-c_\RD$ & $\frac{c_{\MS}-c_{\RD}}{c_{\RD}}$ & $k_\MS-k_\RD$ & $\frac{k_{\MS}-k_{\RD}}{k_{\RD}}$   \\\hline\hline
    CDIMA I & $D_2 =\mu$, $a=12,~b=0.31,~\mu=50$ & $0.293$ &  $0.092$ & $0.057$  &  $0.006$ & $0.007$  \\ \hline
    CDIMA II &$D_2=2\mu,~a=10.5,~b=0.4,~\mu=13$ & $0.346$ & $0.047$ &  $0.029$ & $-0.022$ & $-0.028$ \\ \hline
    CDIMA III & $D_2 =\mu, ~a=12,~b=0.38,~\mu=50$  & $0.125$ &  $0.035$ & $0.053$  & $-0.006$  & $-0.007$ \\ \hline
    Schnakenberg I & $D_2 =20,~a=0.05,~b=1.4$ & $0.559$ &  $0.087$ & $0.068$ & $-0.016$  & $-0.028$ \\ \hline
    Schnakenberg II & $D_2 =20, ~a=0.13,~b=1.4$ & $0.241$  & $0.062$ & $0.131$  & $0.016$  & $0.028$ \\ \hline
  \end{tabular}
\end{center}
\caption{\label{tab1MS} A summary of marginal stability results (index $\MS$) and numerical estimation (index $\RD$) for the key properties of the TW and resultant pattern for CDIMA and Schnakenberg kinetics. We compare each case using relative and absolute error in both the front propagation speed and the wavenumber of the pattern. The domain is one-dimensional with length $L=200$ and $D_1=1$, with $b_{c}$ defined as in Fig \ref{fig2:Sch-Tspace} and Fig \ref{figA1:CDIMA-Tspace}.}
\end{table}

Next, we  compare the TW velocity prediction over a broader parameter space around the bifurcation point for Schnakenberg kinetics. We use the idea of continuity of solutions of the algebraic marginal stability equations on the bifurcation parameter to estimate the localisation of their solutions for nearby parameter values and then compute the exact solution. This allows calculation autonomy, for example, in the initial estimates of the roots of the algebraic problem (and there might be more than one solution). 
However, the key for comparison is the estimation of the TW speed from the numerical solution of the full RD system, $c_\RD$. When the bifurcation parameter is varied, there is a change in the characteristic time of pattern formation as well as the localisation of the two maxima used to determine the travelling wave velocity. 
Again, to automate the process, we store the locations of the maxima as indicated by the two black vertical lines in the density plots, Fig. \ref{fig1:Sch-PDEsolu}, and use the continuous dependence of the solution on the parameter to find their localisation after a small variation of the bifurcation parameter in their neighbourhood.  
Finally, we use 10k points for spatial discretisation (further refinement does not lead to more than $1\%$ difference in the values of the final pattern). 

\begin{figure}
\centering
\begin{subfigure}{.48\textwidth}
    \centering
    \includegraphics[width=.95\linewidth]{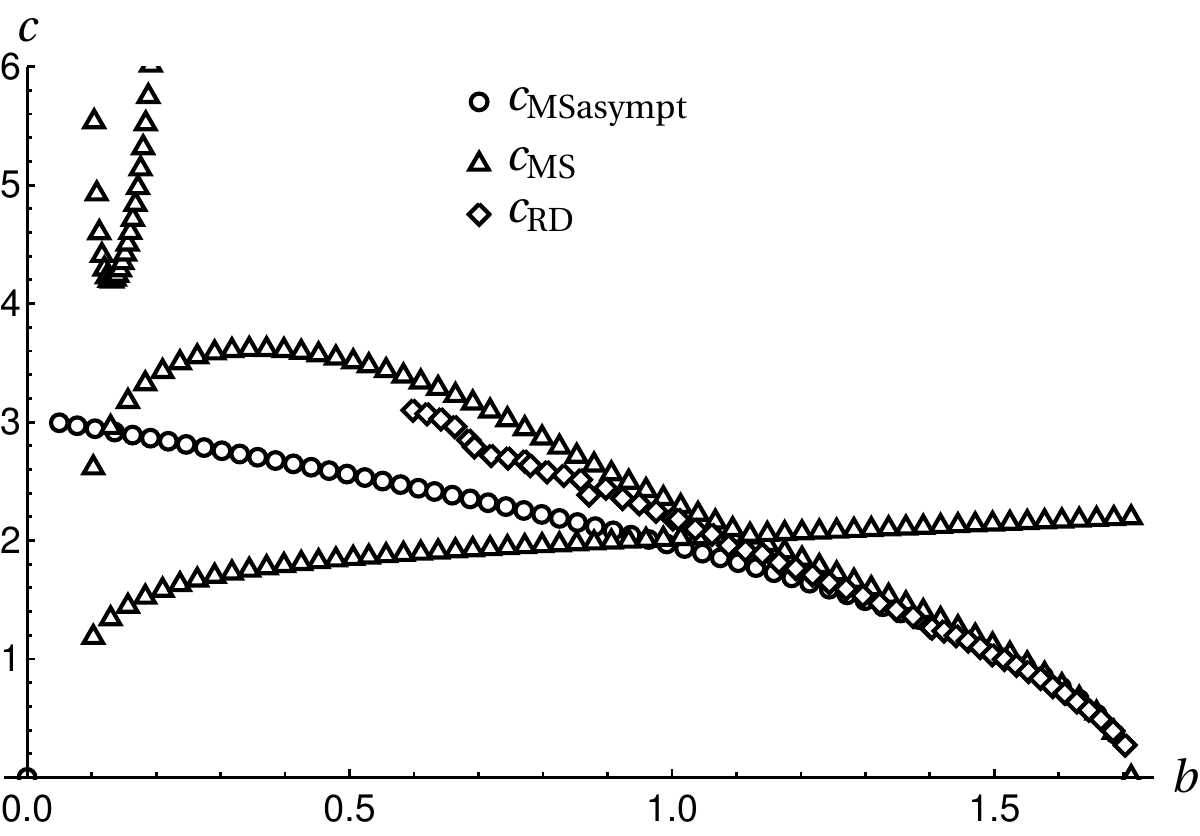}
    \caption{Convergence of all the methods to the same square root behaviour close to the bifurcation point $b_c=1.712$ is apparent. Note that the marginal stability approach shows multiple roots (multiple values of predicted TW speeds) as there are three branches of $c_\MS$, two of them existing near the bifurcation point.}
    \label{fig:BifDiagSch}
  \end{subfigure}
  \hspace{0.3cm}
\begin{subfigure}{.48\textwidth}
    \centering
    \includegraphics[width=.95\linewidth]{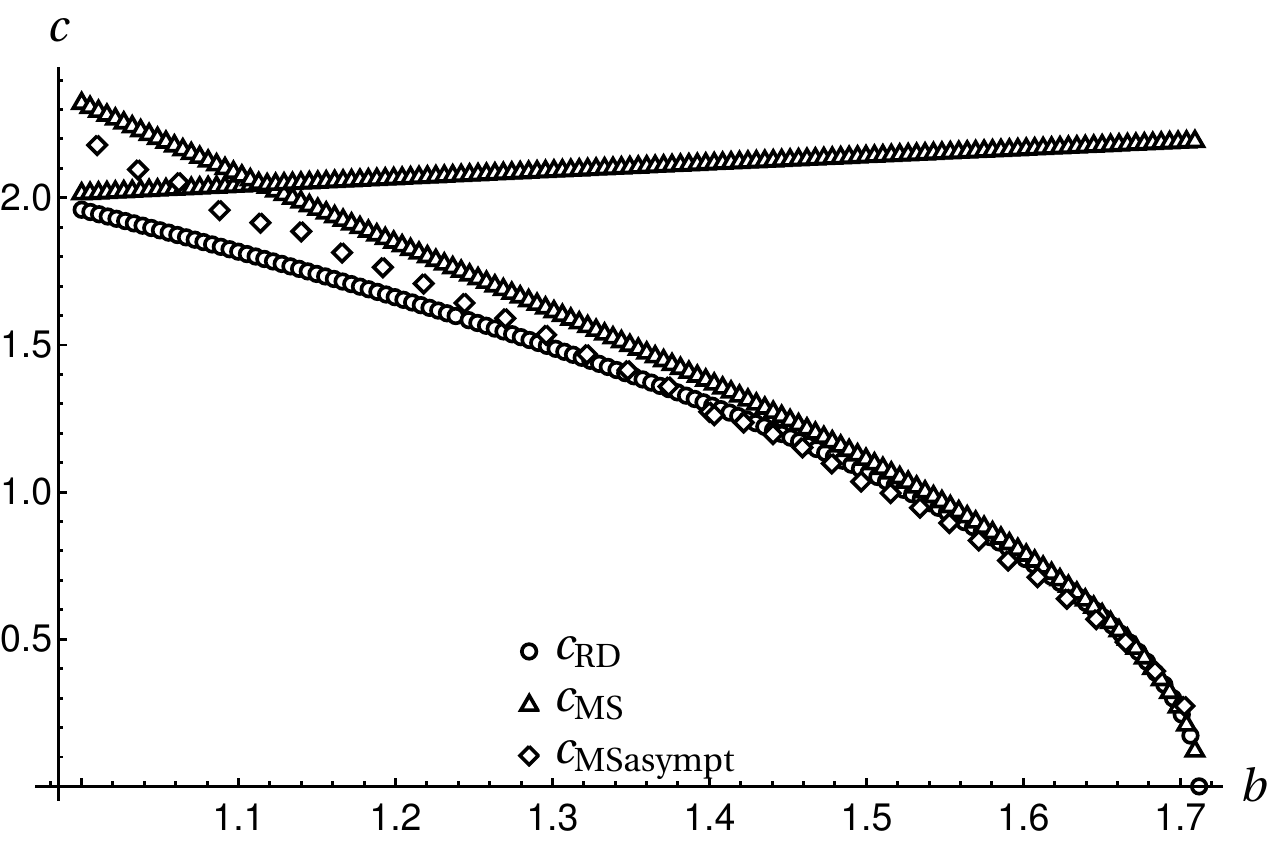}
    \caption{A magnification of the figure in the left panel closer to the bifurcation point showing the convergence of the marginal stability asymptotics to the marginal stability criterion results.}
    \label{fig:BifDiagSchzoom}
\end{subfigure}
\caption{Travelling wave speed (given in $x,~t$ units) as predicted from (i) the full solution of the RD problem with Schnakenberg kinetics, $c_\RD$; (ii) from the marginal stability conditions, $c_\MS$; and (iii) from the marginal stability asymptotics, $c_{\rm MSasympt}$. Parameter values were chosen as $D_1 =1,~ D_2 =20$, $a=0.05$, and 1D domain with size $L=200$ (Schnakenberg I). Note that the choice of the magnitude of the localised initial condition in the finite domain is important, as demonstrated in Fig \ref{fig3:Sch-CvsIC} and we consider $\rho=3$. Note that the marginal stability approach can yield multiple solutions, as illustrated here for the whole range of parameter $b$. Finally, the $c_{\RD}$ predictions were only calculated from the bifurcation point $b_{c}$ to $b\approx 0.6$ due to finite size effects such as the discretely decreasing number of maxima used to estimate $c_{\RD}$ and the longer duration of transient effects before the pattern stabilises.} 
\label{fig4:Cvsb}
\end{figure}

The results of this approach are shown in Fig \ref{fig4:Cvsb} with plots for the TW velocity in $x,~t$ variables. It can be seen that we indeed observe a behaviour corresponding to the square root of the distance from the bifurcation point in its neighbourhood and that there is a match between the asymptotic expression for the travelling wave speed close to the bifurcation point and the numerical solution of the marginal stability criterion. 
Finally, note that the equations for marginal stability may have multiple solutions, as can be seen in Fig \ref{fig4:Cvsb}, where multiple branches of $c_\MS$ appear (even close to the bifurcation point). 


\subsection{Envelope equation, GLE}
We now consider the amplitude equation approach. We perturb the HSS $\bm{v}=\bm{u}-\bm{u}^*$ and expand to identify the approximate evolution equation of the perturbation:
\begin{equation}
  \label{eq:pert}
  \frac{\partial \bm{v}}{\partial t} = \LI {\bf v}+ \LII {\bf v}{\bf v}+\LIII {\bf v}{\bf v}{\bf v}, 
\end{equation}
where the linear operator $\LI={\bf D}\frac{\partial^2 }{\partial x^2}+{\bf J}$ is the linearised RD problem with $J_{ij}=R_{i,u_j}$ evaluated at ${\bf u}^*$, which is the case for higher derivatives below as well. 
As we assume smooth kinetics $\bm{R}$, $\LII$ and $\LIII$ are symmetric and 
we have denoted the nonlinear terms in the following way
\begin{align*}
  (\LII{\bf v}{\bf v})_i&=\frac{1}{2} \sum_{j,k}R_{i,u_ju_k} v_j v_k\\
  (\LIII{\bf v}{\bf v}{\bf v})_i&=\frac{1}{6} \sum_{j,k,l}R_{i,u_ju_ku_l} v_j v_k v_l,
\end{align*}
hence  $\LII$ represents the quadratic and $\LIII$ the cubic kinetics approximation.

Now, recalling the bifurcation parameter in general is $\alpha=\epsilon^{2}$, 
the regular expansion of the perturbation near the bifurcation point $\alpha=0$ 
\begin{equation*}
  \bm{v}=0+\epsilon \bm{v}_1+\epsilon^2 \bm{v}_2+\epsilon^3 \bm{v}_3+\ldots,
\end{equation*}
gives
\begin{align*}
  {\bf J}&={\bf J}_0+\epsilon^2 {\bf J}_1+\epsilon^4{\bf J}_2+\ldots,\\
  \LII&=\LII_0+\epsilon^2 \LII_1+\epsilon^4\LII_2+\ldots,\\
  \LIII&=\LIII_0+\epsilon^2 \LIII_1+\epsilon^4\LIII_2+\ldots.\\
\end{align*}
We shall denote $\LI_j={\bf D}\frac{\partial^2 }{\partial x^2}+{\bf J}_j$. 

Taking advantage of the known relation in scaling of time and space near a bifurcation point \citep{hoyle2006pattern} to balance two spatial derivatives with the one temporal derivative, we introduce the following set of \emph{slow} variables:
\begin{equation*}
   \tau=\epsilon^2 t, \quad y=\epsilon x,
 \end{equation*}
 while we consider the original independent variables $t,~x$ as \emph{fast}. Using the method of multiple scales, writing $v_j$ as functions of both slow and fast variables, that is $v_j(t,\tau,x,y)$, where
   \begin{equation*}
 \frac{\partial}{\partial t} \to \frac{\partial}{\partial t} + \epsilon^2 \frac{\partial}{\partial \tau}, \quad \frac{\partial}{\partial x} \to \frac{\partial}{\partial x} + \epsilon \frac{\partial}{\partial y}
\end{equation*}
and by equating powers of $\epsilon$, we get:
\begin{subequations}
\begin{equation}
  \label{eq.Oeps1}
  \epsilon^1:\quad \frac{\partial}{\partial t} \bm{v}_1 - \LI_0 \bm{v}_1=0,
\end{equation}
\begin{equation}
  \label{eq.Oeps2}
  \epsilon^2:\quad \frac{\partial}{\partial t} \bm{v}_2 - \LI_0 \bm{v}_2=2 {\bf D} \frac{\partial^2}{\partial x\partial y} \bm{v}_1+\LII_0 \bm{v}_1\bm{v}_1,
\end{equation}
\begin{equation}
  \label{eq.Oeps3}
  \epsilon^3:\quad \frac{\partial}{\partial t} \bm{v}_3 - \LI_0 \bm{v}_3=-\frac{\partial}{\partial \tau} \bm{v}_1+2 {\bf D} \frac{\partial^2}{\partial x\partial y} \bm{v}_2 +{\bf D} \frac{\partial^2}{\partial y^2} \bm{v}_1+{\bf J}_1 \bm{v}_1+2\LII_0 \bm{v}_1\bm{v}_2+\LIII_0\bm{v}_1\bm{v}_1\bm{v}_1,
\end{equation}
\end{subequations}
while we require strict periodicity in the spatial fast variable, corresponding to the microscale oscillations. 

As we search for an envelope equation for a travelling wave with a stationary profile in $t$ close to a bifurcation point, as motivated in the Introduction, we look for a solution of the form \citep{hoyle2006pattern}
\begin{equation}
  \label{eq:v1Form}
  \bm{v}_1=A(\tau,y) \bm{V} \exp(i k_c x) + c.c.,
\end{equation}
where $k_c$ is the critical wavenumber corresponding to the bifurcation (onset of instability at the bifucation point), and $\bm{V}\exp(i k_c x)$ is the Turing (Fourier) eigenmode of the linearised problem, i.e.
\begin{equation*}
  \LI_0 \bm{V} \exp(i k_c x) = 0,
\end{equation*}
where the explicit expression for $k_c$ is Eq. \eqref{eq:kc}. 
This follows from the requirement that the larger of the two eigenvalues of $-{\bf J_D}$, where ${\bf J_D}=k_c^2 {\bf D}-{\bf J}_0$, is zero
. Note that $\bm{V}$ is the (right) eigenvector of ${\bf J_D}$ associated with the zero eigenvalue. 
We remark that this assumed form of solution, Eq. \eqref{eq:v1Form}, implicitly defines the microscale and hence the periodic boundary conditions are imposed at $(0, 2\pi/k_c)$ in the fast spatial variable $x$. 

By looking at the $\mathcal{O}(\epsilon^2)$ problem \eqref{eq.Oeps2} while taking into account the form of $\bm{v}_1$, see Eq. \eqref{eq:v1Form}, one can see that $\bm{v}_2$ has to be of the form
\begin{equation}
  \label{eq:v2Form}
  \bm{v}_2=\bm{Z}_0(\tau,y) +\bm{Z}_1(\tau,y) \exp(i k_c x)+\bm{Z}_2(\tau,y) \exp(2i k_c x) + c.c.,
\end{equation}
where $c.c.$ denotes complex conjugate. In particular, by comparing the coefficients of eigenvectors (invoking their orthogonality, see below), we obtain
\begin{align*}
  1:&\quad -({\bf J}_0\bm{Z}_0)_j = 2 |A|^2 (\LII_0  \bm{V} \bar{\bm{V}})_j,\\
  \exp(i k_c x):&\quad k_c^2 ({\bf D} \bm{Z}_1)_j-({\bf J}_0 \bm{Z}_1)_j= i k_c 2 \frac{\partial A}{\partial y} ({\bf D} \bm{V})_j,\\
  \exp(2 i k_c x):&\quad 4 k_c^2 ({\bf D} \bm{Z}_2)_j-({\bf J}_0 \bm{Z}_2)_j= A^2 (\LII_0 \bm{V} \bm{V})_j,
\end{align*}
noting that the relations for the complex conjugate are exactly the same. From the first and last equations we directly have:
\begin{subequations}
  \begin{equation}
    \label{eq:Z0}
    \bm{Z}_0=-2|A|^2 {\bf J}^{-1}_0\cdot (\LII_0 \bm{V}\bar{\bm{V}}),
  \end{equation}
  \begin{equation}
    \label{eq:Z2}
    \bm{Z}_2=A^2(4 k_c^2 {\bf D}-{\bf J}_0)^{-1} \cdot(\LII_0 \bm{V}\bm{V}),
  \end{equation}
while for the unknown $\bm{Z}_1$ we obtain the singular equation
  \begin{equation}
    \label{eq:Z1}
    (k_c^2 {\bf D}-{\bf J}_0)\cdot\bm{Z}_1 = 2 i k_c \frac{\partial A}{\partial y} {\bf D}\cdot\bm{V},
  \end{equation}
  as $(k_c^2 {\bf D}-{\bf J}_0)={\bf J_D}$ and the critical wavenumber $k_c$ is such that ${\bf J_D}$ has a zero eigenvalue. As we shall see below, after the discussion of the solvability condition for $\bm{v}_3$, the solvability condition for this singular set of equations is equivalent to the definition of the critical wavenumber $k_c$ and hence $\bm{Z}_1$ is well defined (although not unique).
\end{subequations}

We now return to the second subleading order problem \eqref{eq.Oeps3}. As its left-hand side is the same as the leading order problem which has a solution for zero right-hand side, we know from the Fredholm alternative (recalling that we assume $\bm{v}_j$ is independent of the fast variable $t$) that the right-hand side of Eq. \eqref{eq.Oeps3} has to be perpendicular to the solution $\bm{f}$ of the adjoint homogeneous problem $(\LI_0)^* f=0$.

From the definition of the adjoint operator, $\langle \LI_0 f,g\rangle = \langle f,(\LI_0)^* g\rangle $, and  recalling the strict periodicity on $(0,2\pi/k_c)$, we have 
\begin{equation*}
  \langle \LI_0 f,g\rangle = \int_0^{2 \pi/k_c} (\LI_0 f)^T \bar{g} dx=\int_0^{2 \pi/k_c} \left( \frac{\partial^2 \bm{f}^T}{\partial x^2}\cdot{\bf D} + ({\bf J}_0\bm{f})^T\right)  \bar{\bm{g}} dx=\int_0^{2 \pi/k_c} \bm{f}^T\left({\bf D} \frac{\partial^2 \bm{\bar{g}}}{\partial x^2}+{\bf J}_0^T\bar{\bm{g}}\right) dx,
\end{equation*}
and hence $(\LI_0)^*\bm{g} = {\bf D} \frac{\partial^2 \bm{g}}{\partial x^2}+{\bf J}_0^T\bm{g}$.

Therefore, the solvability condition is the requirement of perpendicularity to the solution to $(\LI_0)^* \bm{g}=0$. We claim that such solutions are the left eigenvectors of ${\bf J_D}$, i.e. $\bm{g}=A(\tau,y) \bm{W} \exp(i k_c x) +c.c.$, where, WLOG, $\bm{W}^T\cdot\bm{V}=1$ (considering simple eigenvalues of ${\bf J}$).

Choosing $\bm{V}=(1,v)^T$ and $\bm{W}=\frac{1}{1+vw}(1,w)^T$, we have $\bm{W}^T\cdot\bm{V}=1$, while for $\bm{V}$ to be the (right) eigenvector of ${\bf J_D}$ corresponding to the zero eigenvalue we require
\begin{equation}
  \label{eq:v}
  v=-\frac{1}{(J_0)_{12}}((J_0)_{11}-k_c^2 D_1)=-\frac{(J_0)_{21}}{((J_0)_{22}-k_c^2 D_2)}.  
\end{equation}
Similarly, $w$ follows from the fact that $\bm{W}$ has to be the (right) eigenvector of  ${\bf J_D}^T$ corresponding to the zero eigenvalue, i.e.
\begin{equation}
  \label{eq:w}
  w=-\frac{1}{(J_0)_{21}}((J_0)_{11}-k_c^2 D_1)=-\frac{(J_0)_{12}}{((J_0)_{22}-k_c^2 D_2)}.  
\end{equation}

Hence, we may now write the first order correction $\bm{v}_2$  \eqref{eq:v2Form} explicitly as 
\begin{subequations}
  \begin{equation}
    \label{eq:Z0expl}
    \bm{Z}_0=-2|A|^2 \frac{1}{\det {\bf J}_0} \begin{pmatrix}(J_0)_{22} r_1 -(J_0)_{12} r_2\\-(J_0)_{21} r_1 + (J_0)_{11} r_2\end{pmatrix},
  \end{equation}
  where we set $r_j=\frac{1}{2}\left(R_{j,v_1v_1}+2vR_{j,v_1v_2}+v^2R_{j,v_2v_2}\right)$. 
  Furthermore,
  \begin{equation}
    \label{eq:Z2expl}
    \bm{Z}_2=-A^2\frac{1}{\det (4 k_c^2 {\bf D}-{\bf J}_0)} \begin{pmatrix}(4 k_c^2 D-J_0)_{22} r_1 -(4 k_c^2 D-J_0)_{12} r_2\\-(4 k_c^2 - J_0)_{21} r_1+ (4 k_c^2 D-J_0)_{11} r_2\end{pmatrix},
  \end{equation}
  \begin{equation}
    \label{eq:Z1expl}
    \bm{Z}_1=\begin{pmatrix}z\\v z - 2 i k_c \frac{\partial A}{\partial y} \frac{D_1}{(J_0)_{12}}\end{pmatrix},
  \end{equation}
  where $z$ is arbitrary and the solvability condition for $\bm{Z}_1$, eq. \eqref{eq:Z1}, reads $D_1+v w D_2=0$ (being equivalent to the definition of $k_c$).  Note that the arbitrariness of $z$, and hence non-uniqueness of $\bm{Z}_1$, is manifested by the fact that any multiple of the vector $(1,v)^T=\bm{V}$ can be added to $\bm{Z}_1$. However, this corresponds to the fact that we can add any product $\bm{V}\exp(i k_cx)$ to $\bm{v}_2$, i.e. the solution of the equation with zero right-hand side. This part of the solution is already included in the leading order solution, i.e. in the solution of the problem \eqref{eq.Oeps1}, and hence we can set $z=0$ without loss of generality.
\end{subequations}

Finally, the solvability condition for the second subleading problem \eqref{eq.Oeps3} then requires (again, the c.c. terms generate the same expression)
\begin{multline}
  0=\left\langle\frac{\partial}{\partial t} \bm{v}_3 - \LI_0 \bm{v}_3, \bm{W} \exp(i k_c x)\right\rangle\\
  =\int_0^{2\pi/k_c} \overline{\exp(i k_c x)}\bm{W}^T. \left[-\frac{\partial}{\partial \tau} \bm{v}_1+2 {\bf D}\cdot \frac{\partial^2}{\partial x\partial y} \bm{v}_2 +{\bf D}\cdot \frac{\partial^2}{\partial y^2} \bm{v}_1+{\bf J}_1 \cdot\bm{v}_1+2\LII_0 \bm{v}_1\bm{v}_2+\LIII_0\bm{v}_1\bm{v}_1\bm{v}_1\right]dx=\\
  =\bm{W}^T\cdot\left[-\frac{\partial A}{\partial \tau} \bm{V}+{\bf D}\cdot\bm{V} \frac{\partial^2 A}{\partial y^2}+A{\bf J}_1\cdot\bm{V}+2 \LII_0 (A \bm{V}\bm{Z}_0+\bar{A}\bar{\bm{V}} \bm{Z}_2)+3 A |A|^2 \LIII_0 \bm{V}\bar{\bm{V}}\bm{V}+2{\bf D}\cdot \begin{pmatrix}0\\2 k_c^2 \frac{D_1}{(J_0)_{12}}\frac{\partial^2 A}{\partial y^2}\end{pmatrix}\right] \\
  =-\frac{\partial A}{\partial \tau}+4 k_c^2 \frac{D_1D_2 w}{(1+vw)(J_0)_{12}} \frac{\partial ^2 A}{\partial y^2}+A \bm{W}^T\cdot{\bf J}_1\cdot \bm{V}+A|A|^2 \bm{W}^T\cdot\left[2\LII_0(\bm{V}\bm{Z}_0/|A|^2)+3\LIII_0(\bm{V}\bm{V}\bm{V})+2\LII_0(\bm{V}\bm{Z}_2/A^2) \right]
\end{multline}
noting that $\bm{W}^T\cdot{\bf D}\cdot\bm{V}=0$ from the solvability condition of $\bm{Z}_1$.

Therefore the scalar envelope equation is the real Ginzburg-Landau equation
\begin{equation}
  \label{eq:AmplitudeGLE}
  \frac{\partial A}{\partial \tau}=d \frac{\partial ^2 A}{\partial y^2}+c_1 A -c_3 A |A|^2,
\end{equation}
where $d=4 k_c^2 \frac{D_1D_2 w}{(1+vw)(J_0)_{12}}$ and the coefficient of the linear term is
\begin{equation*}
  c_1= \bm{W}^T\cdot{\bf J}_1\cdot \bm{V}=\frac{1}{1+vw}\left((J_1)_{11}+v (J_1)_{12}+w (J_1)_{21}+vw (J_1)_{22}\right),
\end{equation*}
while the cubic term is
\begin{align*}
c_3&=-\bm{W}^T\cdot\left[2\LII_0(\bm{V}\bm{Z}_0/|A|^2)+3\LIII_0(\bm{V}\bm{V}\bm{V})+2\LII_0(\bm{V}\bm{Z}_2/A^2) \right]\\
     &= - \frac{1}{1+vw}\Bigg\{-\frac{2}{\det {\bf J}_0}\Bigg[r_1\Bigg((J_0)_{22}\left(R_{1,v_1v_1}+v R_{1,v_2v_1}+wR_{2,v_1v_1}+vw R_{2,v_2v_1}\right)\\
     &\hspace{4cm}-(J_0)_{21}\left(R_{1,v_1v_2}+v R_{1,v_2v_2}+wR_{2,v_1v_2}+vw R_{2,v_2v_2}\right)\Bigg)+\\
   &\hspace{2cm} r_2\Bigg((J_0)_{11}\left(R_{1,v_1v_2}+v R_{1,v_2v_2}+wR_{2,v_1v_2}+vw R_{2,v_2v_2}\right)\\ &\hspace{4cm}-(J_0)_{12}\left(R_{1,v_1v_1}+v R_{1,v_2v_1}+wR_{2,v_1v_1}+vw R_{2,v_2v_1}\right)\Bigg)\Bigg]+\\
     &+\frac{1}{2}\Big[R_{1,v_1v_1v_1}+3vR_{1,v_1v_1v_2}+3v^2R_{1,v_1v_2v_2}+v^3R_{1,v_2v_2v_2}\\
 & \hspace{2cm}+w\left(R_{2,v_1v_1v_1}+3vR_{2,v_1v_1v_2}+3v^2R_{2,v_1v_2v_2}+v^3R_{2,v_2v_2v_2}
    \right)\Big]\\
     &   +\frac{1}{\det({\bf J}_0-4 k_c^2{\bf D})}\Bigg[\left(\left[(J_0)_{11}-4k_c^2 D_1\right]r_2-(J_0)_{21}r_1\right)\left(R_{1,v_1v_2}+v R_{1,v_2v_2}+wR_{2,v_1v_2}+vw R_{2,v_2v_2}\right)\\
  &\hspace{2cm}+\left(\left[(J_0)_{22}-4k_c^2 D_2\right]r_1-(J_0)_{12}r_2\right)\left(R_{1,v_1v_1}+v R_{1,v_2v_1}+wR_{2,v_1v_1}+vw R_{2,v_2v_1}\right)\Bigg]\Bigg\}.
\end{align*}

\subsection{Travelling wave analysis}

First, let us rewrite the amplitude equation \eqref{eq:AmplitudeGLE} in a more suitable form for the travelling wave analysis.
Note that in our studied case, we know that $c_1>0$ because we are considering a primed Turing system where a travelling wave develops due to a localised perturbation. Specifically, we are concerned with the system with parameter values in Turing space and hence the homogeneous steady state solution corresponding to $A=0$ is linearly unstable, i.e. $c_1>0$. 
Similarly, we naturally expect the existence of a fixed profile of the desired travelling wave, hence the existence of a positive stationary and homogeneous root of the amplitude equation $A^*>0$. 
Thus we have $c_{3}>0$ to allow for such a solution. 
We now rewrite the amplitude equation \eqref{eq:AmplitudeGLE} as
\begin{equation}
  \label{eq:AmplitudeA0}
  \frac{\partial A}{\partial \tau}=d \frac{\partial ^2 A}{\partial y^2} -c_3 (A+A^*)A(A-A^*),  
\end{equation}
where $0<A^*=\sqrt{\frac{c_1}{c_3}}$ and we look for a non-negative solution (amplitude).

We introduce the wave variable $\xi=y/\sqrt{d}-\tilde{c}\tau$, consider a particular direction $\tilde{c}>0$, and look for a travelling wave solution with a fixed profile travelling with a speed $c=\tilde{c}\sqrt{d}$. Phase plane analysis to show the existence of the travelling wave is then standard. We rewrite the amplitude equation \eqref{eq:AmplitudeA0} as a first order system in $\xi$
\begin{align*}
  A' &=W,\\
  W' &= -cW + c_3(A+A^*)A(A-A^*),
\end{align*}
where $'$ denotes the derivative with respect to the wave variable $\xi$. The corresponding boundary conditions $(A,W)=(A^*,0)$ at $\xi=-\infty$, $(A,W)=(0,0)$ at $\xi=+\infty$ represent the sought heteroclinic connection between (unstable) $A=0$ and (stable) $A=A^*$.

The fixed point $(0,0)$ is a focus for $\tilde{c}>2\sqrt{c_1}$ and a spiral for $\tilde{c}\in(0,2\sqrt{c_1})$, while $(A^*,0)$ is always a saddle with the unstable manifold in the direction $\left(1,-\frac{\tilde{c}}{2}\left(1-\sqrt{1+8c_3(A^*/\tilde{c})^2}\right)\right)$.

From these observations we can argue that there is a critical speed, $c^*=\tilde{c}^*\sqrt{d}=2\sqrt{c_1d}=2A^*\sqrt{c_3 d}$, below which a travelling wave does not occur. However, there is no unique speed $c$ of a travelling wave as for any $c>c^*$ one can construct a heteroclinic connection between the two fixed points satisfying the desired boundary conditions.

We now use the well known \citet{kolmogorov1937study} results about the asymptotics of the speed of a compactly supported initial condition entailing a travelling wave. Namely, as the cubic kinetics satisfy the requirement of two zeros (which can be scaled to be 0 and 1, respectively), while being positive between them and with the highest derivative at zero, the observed travelling wave speed matches the critical speed (corresponding to the transition from a spiral to a focus) for sufficiently large times.

The analytic profile of the travelling wave corresponding to the critical wave speed is not available. However, we can readily find the profile of a heteroclinic connection for $\tilde{c}=\tilde{c}_A=3\sqrt{\frac{c_1}{2}}=3A^*\sqrt{\frac{c_3}{2}}$ as
\begin{equation}
  \label{eq:AnalyticProfile}
  A_A(\xi) = \frac{A^*}{2}\left(1-\tanh\left(\frac{1}{2}\sqrt{\frac{c_1}{2}}\xi\right)\right)=\frac{A^*}{2}\left(1-\tanh\left(\frac{A^*}{2}\sqrt{\frac{c_3}{2}}\xi\right)\right),
\end{equation}
matching the desired boundary conditions. As $\tilde{c}^*=\tilde{c}_A\frac{2\sqrt{2}}{3}= \tilde{c}_A(1-\delta)$ with $\delta\approx 0.057$, we can rewrite the equation for the travelling wave profile as
\begin{equation} \label{TWprofileDelta}
    0 = \tilde{c}_A(1-\delta)A' + A'' - c_3(A+A^*)A(A-A^*),
\end{equation}
and hence expect the profile to be closely represented by the analytic profile $A_A(\xi)$, eq. \eqref{eq:AnalyticProfile}, for $\delta=0$.

It can be shown that the appropriate form of the asymptotic expansion for $A$ is
\begin{equation*}
  A=A_0+\delta^{1/2}A_1+\delta A_2+\mathcal{O}(\delta^{3/2}),
\end{equation*}
and collecting powers of $\delta$ after substituting this into Eq. \eqref{TWprofileDelta} gives $A_0(\xi)=A_A(\xi)$ and
\begin{align*}
  \delta^{1/2}:\quad 0=\tilde{c}_A A_1'+A_1''-c_3A_1(3A_0^2-(A^*)^2),\\
  \delta:\quad 3 c_3 A_1^2 A_0+\tilde{c}_AA_0'=\tilde{c}_A A_2'+A_2''-c_3A_2(3A_0^2-(A^*)^2).
\end{align*}
As the solution to the first equation is $A_1(\xi)=K A_0'(\xi)$, which satisfies the required homogeneous Dirichlet boundary conditions at $\pm\infty$, the solvability condition for the second equation is
\begin{equation*}
 0= \int_R \left(3 c_3 A_1^2(\xi) A_0(\xi)+\tilde{c}_A A_0'(\xi)\right) A_0'(\xi) d\xi.
\end{equation*}
Using the expected form of solution $A_1=K A_0'$, and invoking the exact form of $A_0(\xi)$, Eq. \eqref{eq:AnalyticProfile}, a direct calculation yields
\begin{equation*}
  K=c_1^{-3/2}\frac{10\sqrt{2}}{3}.
\end{equation*}
Finally, the profile of the travelling wave has the following analytical approximation
\begin{align}
  A(\xi)&\approx A_A(\xi)+\delta^{1/2} \frac{10 \sqrt{2}}{3}c_1^{-3/2} A_1(\xi)\nonumber\\          &\approx \frac{A^*}{2}\left(1-\tanh\left(\frac{A^*}{2}\sqrt{\frac{c_3}{2}}\xi\right)\right)-\frac{5}{6}\left(1-\frac{2\sqrt{2}}{3}\right)^{1/2}\frac{(A^*)^{1/2}}{c_3^{1/4}} \frac{1}{\cosh^2\left(\frac{A^*}{2}\sqrt{\frac{c_3}{2}}\xi\right)}.\label{eq:TWprofileApprox}
\end{align}
Note that this profile corresponds to a heteroclinic connection for a cubic reaction kinetics between an unstable node and a saddle that is unique in the above sense due to Kolmogorov's result.

In short, the concept of a characteristic speed of pattern propagation in a reaction-diffusion system is well founded (thanks to the dominance of the critical wave speed) and corresponds 
 to 
\begin{equation}
  \label{eq:NatSpeedOfPattern}
c_{\rm{env}}=\epsilon c = \epsilon c^*=2\epsilon\sqrt{c_1 d}= 2 \epsilon \sqrt{\frac{d}{1+vw}\left((J_1)_{11}+v (J_1)_{12}+w (J_1)_{21}+vw (J_1)_{22}\right)},
\end{equation}
in the original $x,~t$ dimensional variables.   
We recall that $d=4 k_c^2 \frac{D_1D_2 w}{(1+vw)(J_0)_{12}}$, $v=-\frac{(J_0)_{21}}{((J_0)_{22}-k_c^2 D_2)}$ and $w=-\frac{(J_0)_{12}}{((J_0)_{22}-k_c^2 D_2)}$, see Eqns. \eqref{eq:v}, \eqref{eq:w}. The travelling wave profile can be approximated by Eq. \eqref{eq:TWprofileApprox}.

Crucially, it can be shown that the expression for the TW speed obtained from the envelope equation, Eq. \eqref{eq:NatSpeedOfPattern}, and from the marginal stability conditions close to the bifurcation point, Eq. \eqref{eq.cMSBifP}, are exactly equivalent, when we take into account the conditions that hold at the bifurcation point \eqref{eq:bifPointCond} and $D_1+v w D_2=0$ (being equivalent to the definition of $k_c$). 
To see this, let us compare the ratios of the coefficients $\frac{(J_1)_{12}}{(J_1)_{11}}$, $\frac{(J_1)_{21}}{(J_1)_{11}}$ and $\frac{(J_1)_{22}}{(J_1)_{11}}$ in the squared expression for the TW velocity near a bifurcation point in the marginal stability approach, Eq. \eqref{eq.cMSBifP}. 
If we show that they are in the ratio $v,~w$ and $v w$ then, from Eq. \eqref{eq:NatSpeedOfPattern}, we will be left with proving that the coefficients of $(J_1)_{11}$ match in both expressions. First, the ratio of the coefficients of $(J_1)_{22}$ and $(J_1)_{11}$ is $-D_1/D_2$, which is equivalent to $v w$, following from the above expression $D_1=-v w D_2$. 
 Next,  the ratio of the coefficients of $(J_1)_{12}$ and $(J_1)_{11}$ is $\frac{2 D_1 (J_0)_{21}}{D_2 (J_0)_{11}-D_1 (J_0)_{22}}$ which is equivalent to $v=-\frac{(J_0)_{21}}{(J_0)_{22}-k_c^2 D_2}$ via the expression for the critical wavenumber $k_c$. Finally, the ratio of the coefficients of $(J_1)_{21}$ and $(J_1)_{11}$ is $\frac{D_1 (J_0)_{22}-D_2 (J_0)_{11}}{2 D_2 (J_0)_{21}}$, which is equivalent to $w=-\frac{(J_0)_{12}}{(J_0)_{22}-k_c^2 D_2}$ when using the expression for the critical wavenumber $k_c$ and the relation $w=-\frac{D_1}{D_2} v$. Therefore, all that remains is to show that the coefficients of the $(J_1)_{11}$ terms are equal in both expressions \eqref{eq.cMSBifP} and \eqref{eq:NatSpeedOfPattern}. This follows from the fact that the bifurcation point enforces the condition $(D_2 (J_0)_{11}+D_1 (J_0)_{22})^2 = 4 D_1 D_2 \det {\bf J}_0$.

However, in addition to the prediction of the travelling front wave speed and the spatial frequency of the resultant pattern, the amplitude equation approach estimates the amplitude of the final pattern. Hence, in addition to the above numerical verification of the asymptotic version of the marginal stability criterion, we shall assess the estimation of the amplitude of the pattern.

\subsubsection{Assessment of envelope equation performance}

It is worth mentioning that several key characteristics follow from the amplitude equation itself, as estimated  by Eqn. \eqref{eq:TWprofileApprox}. Firstly the asymptotic amplitude of the leading order pattern in $\bm v$ is equal to $2\epsilon$ multiplied by the positive stationary solution of the envelope equation, i.e. $2\epsilon A^*$, with $A^*=\sqrt{c_1/c_3}$, noting the associated component of ${\bf V}=(1,v)$ and the factor of two arising from a complex conjugate,  as follows from Eq. \eqref{eq:v1Form}. We may also compare this leading asymptotic amplitude  directly with the amplitude of the first component $u_1$ of the solution to the full problem.  Second, the speed of the travelling wave is $2\epsilon\sqrt{c_1 d}$ and we have an analytical expression for this in terms of the model parameters, as given by the expressions following Eq. \eqref{eq:AmplitudeGLE}. Third, an estimate of the wave number is immediately available, since the weakly nonlinear theory is based on the localisation of the bifurcation point, which is analytically determinable from linear stability theory, see Eq. \eqref{eq:kc}.


We again consider Schnakenberg and the CDIMA kinetics with the same parameter values as above when testing the marginal stability criterion, hence we consider a primed Turing system. Note that the physical parameter we vary as the bifurcation parameter, denoted $\alpha$, is subject to choice and hence can potentially affect the precision of the analytical results. Nonetheless, the methodology of  deriving the envelope equation is not affected by this choice, though the expansion of the Jacobian given by  $\bm{J}=\bm{J}_0+\alpha\bm{J}_1+\mathcal{O}(\alpha^2)$ generates expressions that ultimately are dependent on the choice of $\alpha$.   Finally, on comparing the full problem with the envelope equation, note that the initial perturbation is of the same magnitude (rescaled by $\epsilon$) and location for both problems, allowing us to use the envelope equation for comparison. However, one distinction is that the perturbation is about a non-trivial homogeneous steady state for the full problem but about the trivial solution for the envelope equation (simply being the zero amplitude of the pattern corresponding to the homogeneous steady state), though this does not invalidate direct comparison between the two solutions.





\begin{figure}
\centering
\begin{subfigure}{.48\textwidth}
    \centering
    \includegraphics[width=.95\linewidth]{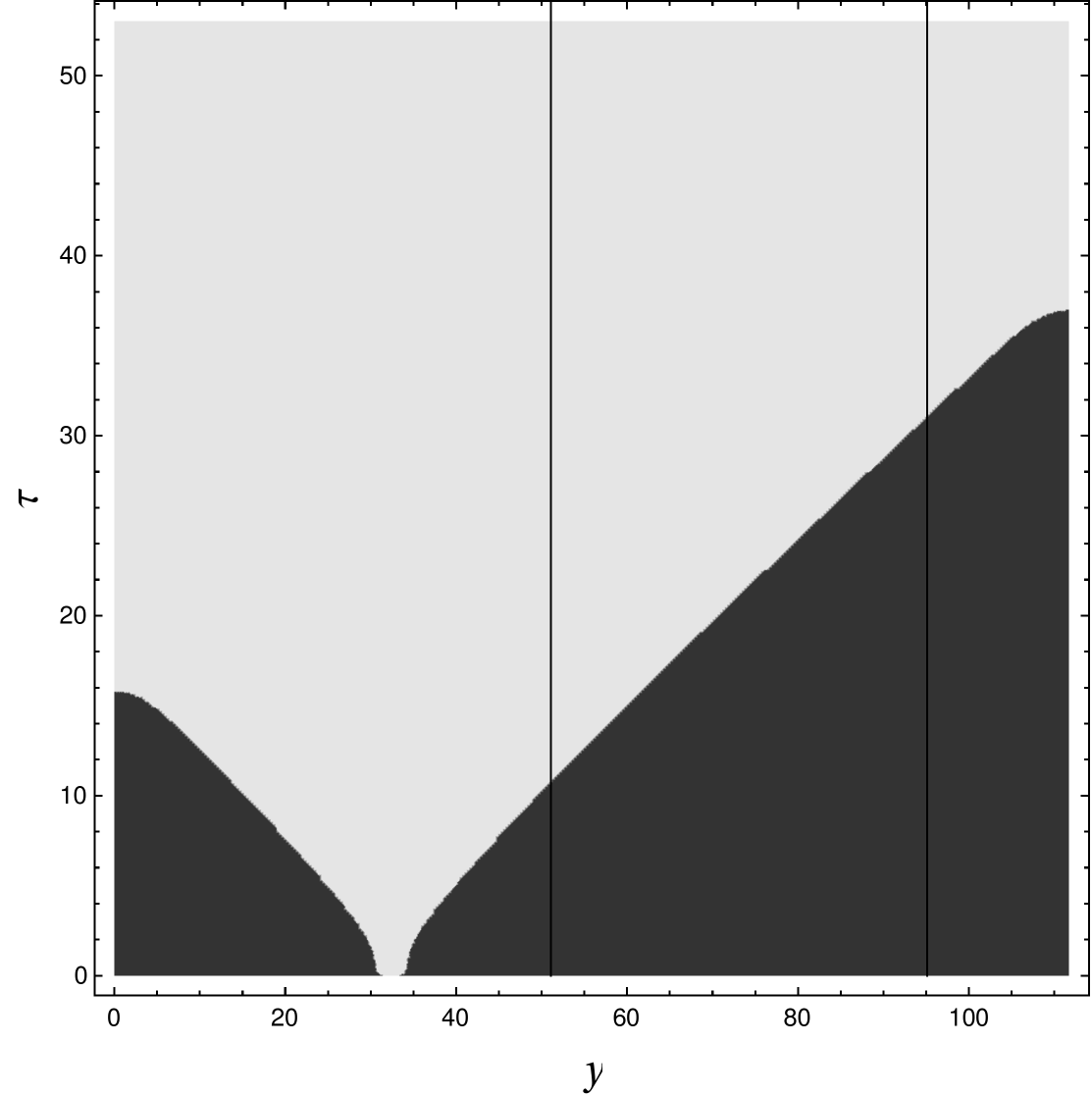}
    \caption{Density plot of the numerical solution to the envelope equation Eq. \eqref{eq.SchenvExample}. The black region is the set of points $(y,\tau)$ where  $A(\tau,y)<A_\env/2$ while the white region shows its complement.}
    \label{fig:Sch-densityEnvelope_halfA}
  \end{subfigure}
  \hspace{0.3cm}
\begin{subfigure}{.48\textwidth}
    \centering
    \includegraphics[width=.95\linewidth]{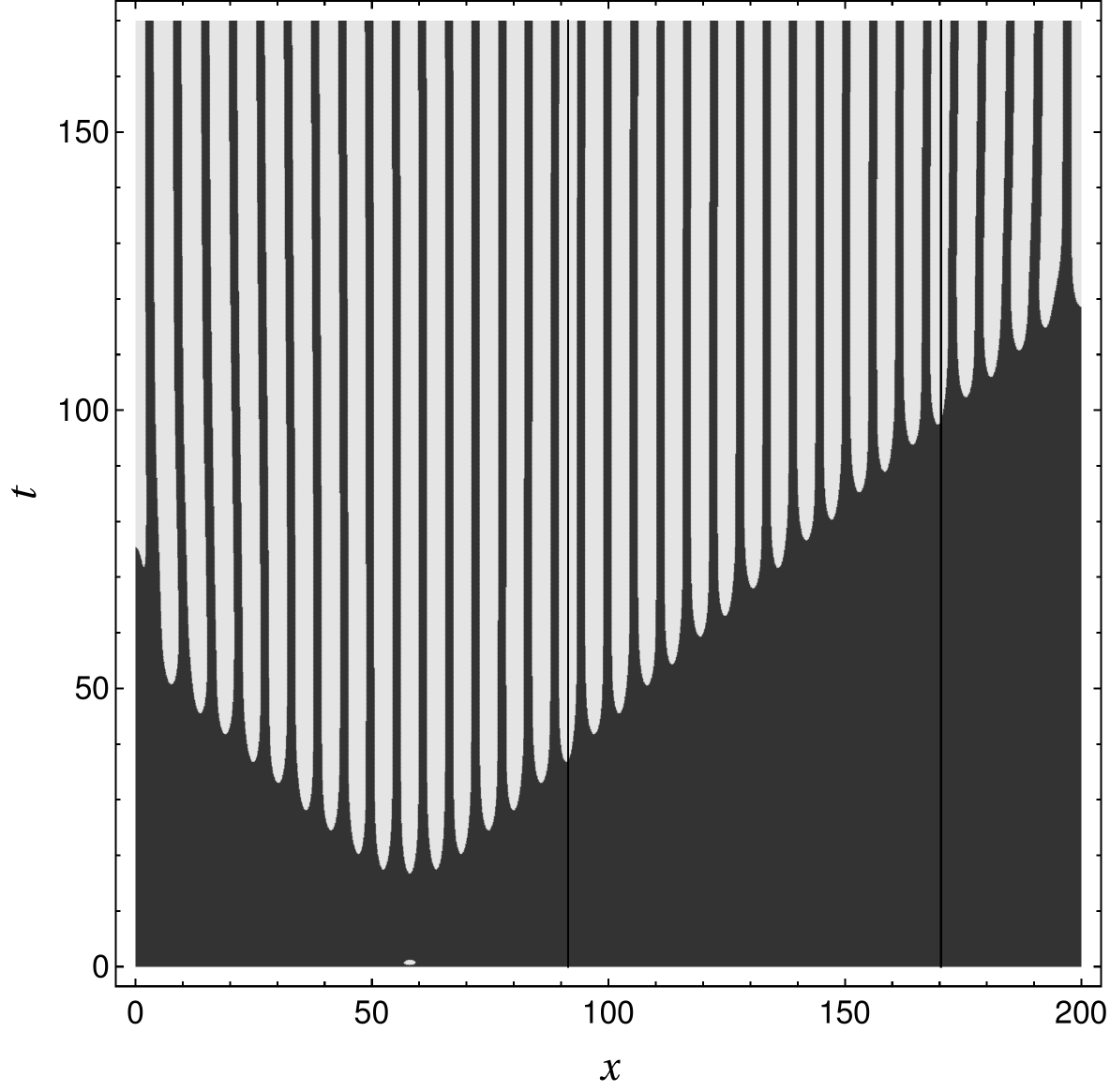}
    \caption{Density plot of $u_1$ of the numerical solution to the full problem.  The black region is the set of points $(x,t)$ where  $A(t,x)<A_\RD/2$ while the white region shows its complement.}
    \label{fig:Sch-densityRD_halfA}
\end{subfigure}
\caption{Schnakenberg kinetics with $D_1=1,~D_2 =20,~a=0.05,~b=1.4$ and a 1D domain with size $L=200$ (Schnakenberg I). The numerical solution for the derived envelope equation is presented in panel (a), and for the full problem, in panel (b). Again, the vertical black lines highlight the positions of cross-sections for determining travelling wave speed and amplitudes. Note the similarity in both the time and location of the pattern initiation and also the development of the spatial pattern behind a front travelling with a fixed velocity (neglecting the initial transition behaviour and boundary effects). We plot the solution to the envelope equation in the corresponding coordinates $\tau=\epsilon^{2}t$, $y=\epsilon x$ (with $\epsilon=0.559$ for the listed parameter values). Hence the two plots are directly comparable and the two thresholded solutions are almost overlapping.}
\label{fig5}
\end{figure}

As a particular example, we consider Schnakenberg kinetics with zero flux boundary conditions, the parameter values $D_1=1,~D_2 =20,~a=0.05,~b=1.4$ and a 1D domain with size $L=200$ (Schnakenberg I), see Appendix \ref{App.OtherCases_Envelope} for the other studied cases.
Using the above analysis, we obtain, for this choice of reaction kinetics and parameter values, the following form of the envelope equation
\begin{equation} \label{eq.SchenvExample}
  \frac{\partial A}{\partial \tau}=3.020\frac{\partial^2 A}{\partial y^2} + 0.446 A-0.541 A^3,
\end{equation}
and that $\epsilon=0.559$. Hence, we may simply read out the predicted amplitude of the pattern $A_\env=2\epsilon A^*=1.015$ and using Eq. \eqref{eq:NatSpeedOfPattern} we have an estimate of the travelling wave speed $c_\env=1.297$.  The wavenumber $k_\env$ is equal to the critical wavenumber $k_c$, Eq. \eqref{eq:kc}, and has the value $k_\env=k_c=0.628$. In addition, we solve this scalar reaction diffusion equation with the same zero flux boundary conditions and the same initial condition as above in the full problem, that is, a localised perturbation of the same magnitude, $\rho=3$, and location $x=60$ about the unstable trivial homogeneous steady state. 
Plotting this solution reveals a good match with the solution to the full problem, see Figures \ref{fig:Sch-densityEnvelope_halfA}, \ref{fig:Sch-densityRD_halfA}, where we plot the solutions in their natural coordinates.

In addition, from the numerical solution to the envelope equation, we determine the numerically observed speed of the travelling wave $c_\env^\num$ as described above in the two highlighted locations giving $c_\env^\num=1.288$, showing a good match with the critical wave speed determined analytically from the Kolmogorov asymptotic arguments. 

\begin{figure}
\centering
\begin{subfigure}{.47\textwidth}
    \centering
    \includegraphics[width=.95\linewidth]{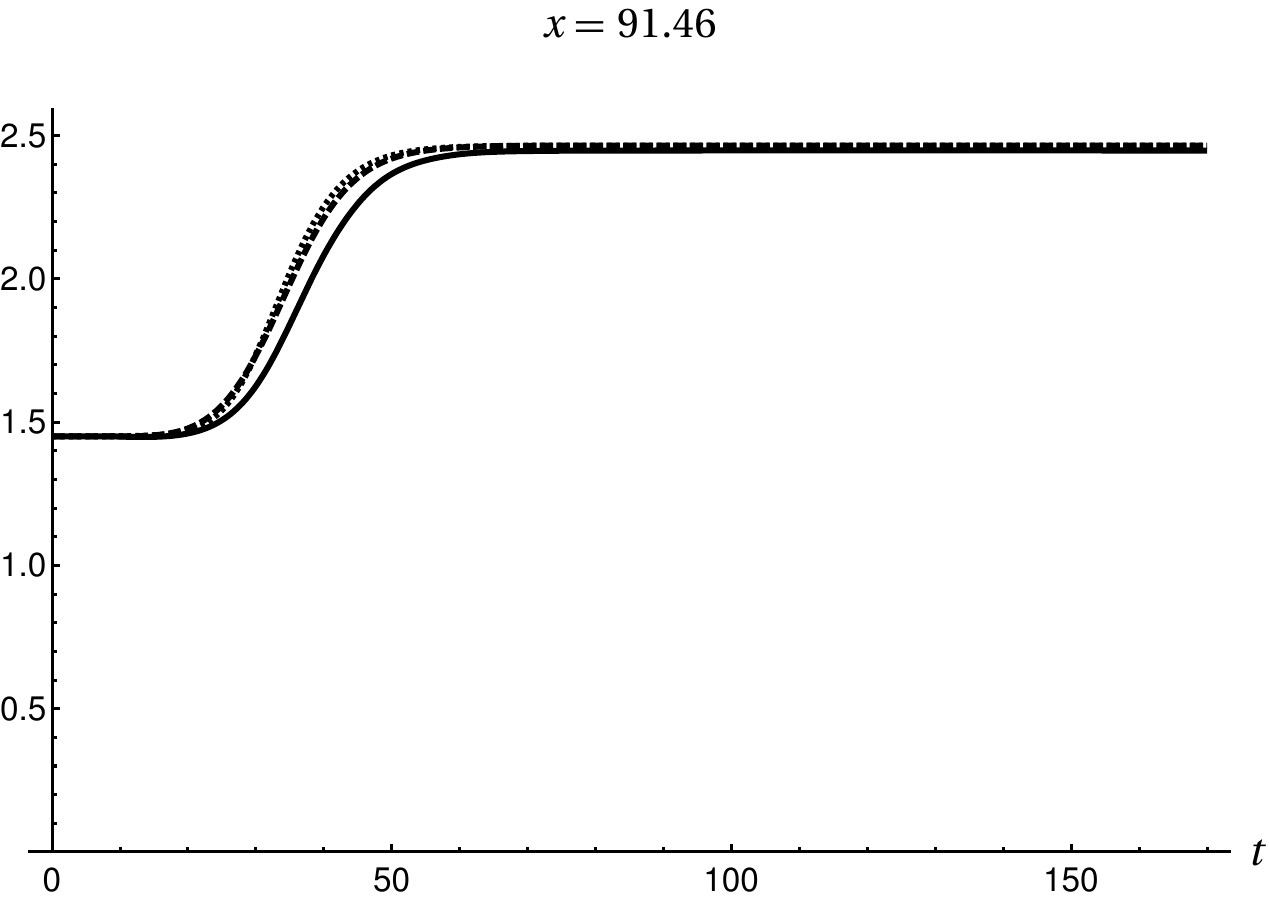}
    \caption{The temporal profile for the numerical solutions and the estimated analytical profile at the first location given by $x=91.46$ in Figs. \ref{fig1:Sch-PDEsolu} and \ref{fig:Sch-densityRD_halfA}. All solutions are shown in the original $t,~x$ variables.}
    \label{fig:Sch-EnvVsPDE-early}
  \end{subfigure}
  \hspace{0.5cm}
\begin{subfigure}{.47\textwidth}
    \centering
    \includegraphics[width=.95\linewidth]{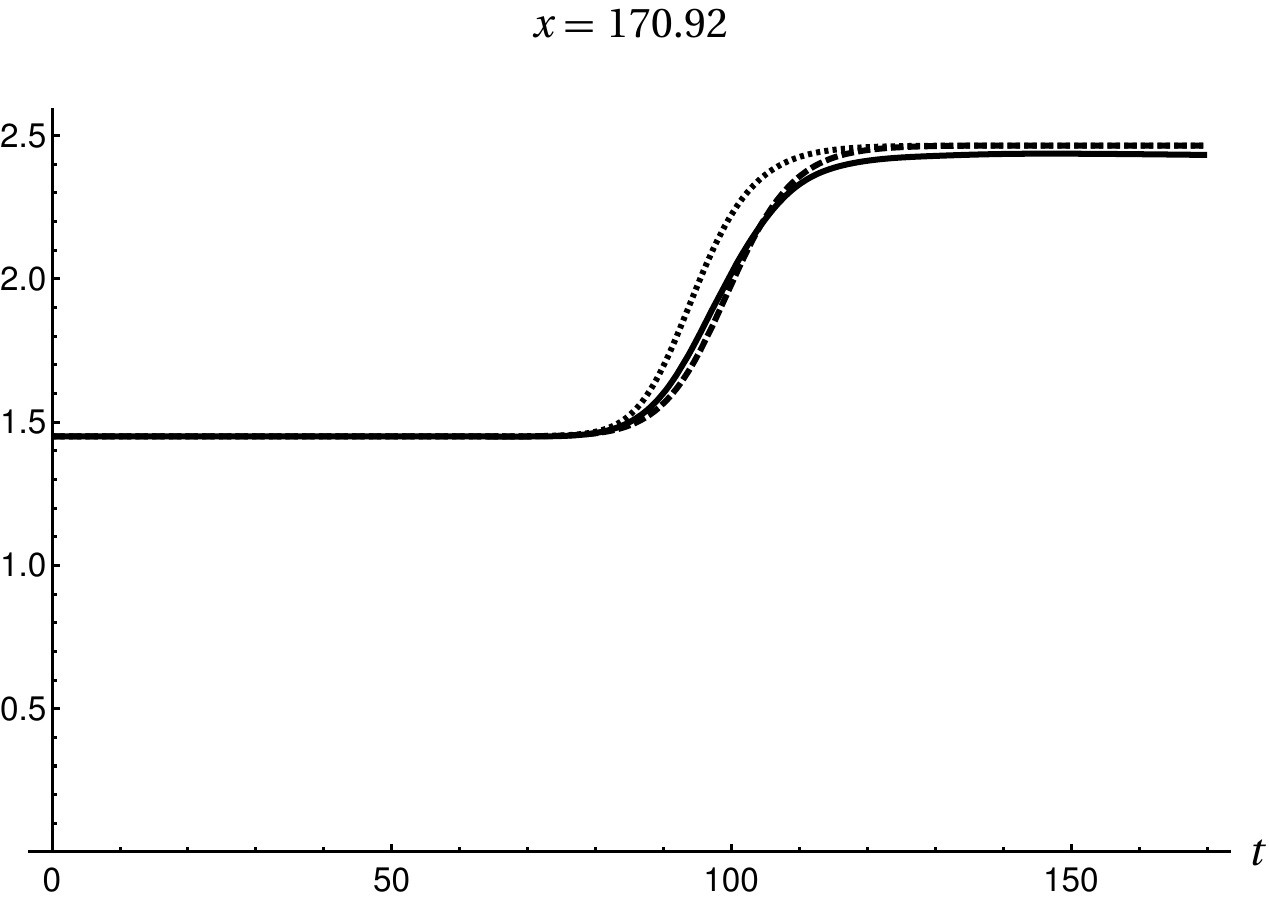}
    \caption{The temporal profile for the numerical solutions and the estimated analytical profile at the second location given by $x=170.92$ in Figs. \ref{fig1:Sch-PDEsolu} and \ref{fig:Sch-densityRD_halfA}. All solutions are shown in the original $t,~x$ variables.}
    \label{fig:Sch-EnvVsPDE-late}
\end{subfigure}
\caption{Schnakenberg kinetics with $D_1=1,~D_2 =20,~a=0.05,~b=1.4$ and a 1D domain with size $L=200$ (Schnakenberg I). Temporal profiles for the solutions at the highlighted positions - vertical black lines in the preceding density plots. The full curve corresponds to the numerical solution $u_1(t,x)$ to the full problem. The dashed curve is the numerical solution to the corresponding envelope problem $(u^*)_1+2\epsilon A(\epsilon^2 t,\epsilon x)$, and the dotted curve (significantly overlapping with the dashed) is the shifted in time estimated analytical profile $(u^*)_1+2\epsilon A\left(\frac{\epsilon}{\sqrt{d}} (x-c_\env \epsilon (t+41))\right)$ from Eq. \eqref{eq:TWprofileApprox}. This shift in the plot of the analytical profile is necessary because it was obtained from phase space in TW coordinates and therefore is subject to the translational invariance of the travelling wave
, see text for more details. As one can observe from the comparison of the two panels, all the three speeds $c_\RD,~c_\env,~c_\env^\num$ of the travelling wave are similar and the approximate analytical profile matches the numerically calculated ones, see the text for more details. Finally, note that there is a disparity in the predicted amplitude of the pattern, the error being $3\%$ of the numerically calculated amplitude from the full problem.}
\label{fig6}
\end{figure}

Finally, we plot the solutions, see Figures \ref{fig:Sch-EnvVsPDE-early},
\ref{fig:Sch-EnvVsPDE-late}, to both the full problem and the envelope equation for the two highlighted cross-sections in Fig. \ref{fig:Sch-densityRD_halfA} corresponding to $x=91.46$ and $x=170.92$.  In this way we can compare the predicted travelling wave profile together with its amplitude and velocity. We also plot the asymptotic estimate of the travelling wave profile, \eqref{eq:TWprofileApprox}, shifted in time by $t_0$ to match the arrival of the wave at the first cross-section with the numerical solution to the envelope equation. 
This time shift $t_0$ is necessary as the determination of the travelling wave profile does not consider initiation and development of the front and thus is free up to a translational shift, which we fix by specifying $t_0$. Its value is determined manually by choosing a value which results in the best match (visually overlapping) between the analytical profile, eq \eqref{eq:TWprofileApprox}, and plotted as a dotted curve, together with a plot of the numerical solution to the envelope problem at the first cross-section (dashed). In particular, we present the function $(u^*)_1+2\epsilon A(\xi)$ where $A(\xi)$ is the identified asymptotic approximation for the amplitude in \eqref{eq:TWprofileApprox} evaluated at a point $\xi(t,x)=\frac{\epsilon}{\sqrt{d}} (x-c_\env \epsilon (t+t_0))$. Therefore, there is a single fitting parameter,  the time shift $t_0$. Therefore, the precision of the analytical estimate of the front velocity $c_\env$ is visually immediate as the difference between the time of arrival of the wave at the second cross-section profile, Fig \ref{fig:Sch-densityRD_halfA}.

We also show the numerical solution to the full RD problem where there is no need to provide a time shift correction. Thus, from the figure we are able to determine the velocities observed in all three approaches: (i) the full reaction diffusion system with numerically estimate speed, $c_\RD$; (ii) the  analytical estimate for the asymptotic travelling wave speed from the envelope equation $c_\env$;  and (iii) the numerically estimated asymptotic travelling wave speed from the numerical solution of the envelope equation,   $c_\env^{\mathrm{num}}$. In addition, we can also assess   the closeness of fit for the front profiles and the time taken for the pattern to develop from the small localised disturbance.

\section{Discussion and conclusions}


The marginal stability criterion was put forward as a conjecture forty years ago \citep{dee1983propagating} and while there are several hypotheses for its derivation \citep{dee1983propagating,ben1985pattern,vansaarloos1988front,tarumi1989wavelength,myerscough1992analysis,chomaz2000propagating,vansaarlos2003front}, for example, being a transition between convective and absolute instabilities \citep{dee1983propagating,rovinsky1992chemical,tobias1998convective,sandstede2000absolute,sherratt2014mathematical,ponedel2017front}, its validity still remains an open problem. In addition, the marginal stability equations may not entail a unique speed of the travelling front, see Fig \ref{fig4:Cvsb}, and do not provide information on the profile or amplitude of the wave.

As an alternative route, we derived the envelope equation for the spatial pattern deposited after the front but it is restricted in validity to be close to the primary bifurcation point. 
However, we were able to show that there is a uniquely selected TW speed via Kolmogorov's argument of dynamical selection via a linear mechanism, also known as the pulled case. 
Note, however, that the selection of TW speed along the lines discussed above, when the TW speed is determined only by what is happening at the front, is limited only to the pulled front case, where its dynamics are driven by linear kinetics around the unstable homogeneous steady state, see \citet{chomaz2000propagating, vansaarloos1988front,berestycki2007generalized}. Pushed fronts require different approaches and will be the subject of future research.



The amplitude equation approach offers several benefits. It not only estimates the front velocity and pattern wavenumber, but also the magnitude of the deposited pattern amplitude and front shape. We derived explicit analytical expressions for all of these properties from the original model parameters. Moreover, the asymptotic solution to the marginal stability criterion yields exactly the same estimate near the bifurcation point. Our analysis establishes a firm basis for the notion of the characteristic speed of pattern propagation in a reaction-diffusion system, while strengthening the motivation for using the marginal stability criterion and provides a footing for the further exploration of a wave of competence \citep{liu2022control}.

In particular, we have shown that the marginal stability criterion and the amplitude equation give exactly the same estimates of the travelling wave speed near primary bifurcation points. Hence, Figure  \ref{fig4:Cvsb} can be also viewed as a comparison of the marginal stability criterion and the amplitude equation approach. This exact match is surprising given the very distinct foundations of the two frameworks but it serves as a confirmation of the marginal stability criterion at least in the vicinity of primary bifurcation points.


Note that other choices of the bifurcation parameter $\alpha$ may yield different predictions as the measure of the distance to the bifurcation  point varies. We leave a more thorough discussion of this effect for future research. Nevertheless, we remark that, for example in the CDIMA kinetics case, the choice $\alpha=(a-a_c)/a_c$, where $a_c=11.058$ is the critical parameter value, yields $\epsilon=0.292$ and hence different predictions of the characteristic features of the travelling wave solutions. Similarly, we also consider the choice $\alpha=(\mu_c-\mu)/\mu_c$, where $\mu_c=63.813$ is the bifurcation point. The associated predictions are listed below in Table \ref{tab2MSenv} and further details including the envelope equations are given in Appendix \ref{App.OtherCases}.

We now summarise the results in the above considered scenarios in Table \ref{tab2MSenv}, restricting the results to relative errors for brevity. Note the role of the choice of the bifurcation parameter while keeping the parameter values the same, lines 1-3 in the table.  
The linear marginal stability criterion showed a good estimation of the TW speed and wavenumber in the studied examples.  Similar results were then obtained for these two properties using a weakly nonlinear analysis and the envelope equation, which we showed to exactly satisfy the marginal stability criterion close to the bifurcation point. Furthermore, while one might expect a loss of accuracy of the amplitude equation approach with increasing distance from the bifurcation point, the marginal stability condition is not restricted to the neighbourhood of the bifurcation point and shows good estimates even outside this neighbourhood. However, as we have shown, there is the  problem of multiple possible solutions, among which the correct solution corresponding to the characteristic TW speed cannot be selected without further analysis or insight. The envelope equation approach predicted that the evolution of the travelling wave in the envelope equation is slightly delayed compared to the actual full reaction-diffusion problem, the amplitude approach allows us to capture, qualitatively, the shape of the solution and, in many cases, is also reasonably accurate quantitatively, see Table \ref{tab2MSenv}.

\begin{table}
  \begin{center}
  \begin{tabular}{ | l | c | c | c | c | c | c | c |}
    \hline
    & Parameters   & $\epsilon= \alpha^{1/2}$ & $\frac{c_{\RD}-c_\MS}{c_\RD}$ & $\frac{c_{\RD}-c_\env}{c_\RD}$ & $\frac{k_{\RD}-k_\MS}{k_\RD}$ & $\frac{k_{\RD}-k_\env}{k_\RD}$ & $\frac{A_{\RD}-A_\env}{A_\RD}$  \\\hline\hline
    CDIMA I & $D_2 =\mu$, $a=12,~b=0.31,~\mu=50$ & $\sqrt{b_c-b}=0.293$ &   -0.057 & 0.010  & -0.007 & -0.096 & 0.529 \\ \hline
    CDIMA I & same as above & $\sqrt{\frac{a-a_c}{a_c}}=0.292$ &   -0.057 & -0.222  & -0.007 & -0.107  & 0.678 \\ \hline
    CDIMA I & same as above & $\sqrt{\frac{\mu_c-\mu}{\mu_c}}=0.465$ &  -0.057 & 0.010  & -0.007 &  -0.096 & 0.529 \\ \hline
    CDIMA II &$D_2=2\mu,~a=10.5,~b=0.4,~\mu=13$ & $\sqrt{b_c-b}=0.346$ &  -0.029 & 0.040 & 0.028 & -0.074 & 0.132 \\ \hline
    CDIMA  III & $D_2 =\mu, ~a=12,~b=0.38,~\mu=50$ & $\sqrt{b_c-b}=0.125$ & -0.053 & -0.042  & 0.007 & -0.007 & 0.217 \\ \hline
    Schnakenberg I & $D_2 =20,~a=0.05,~b=1.4$ & $\sqrt{b_c-b}=0.559$ &  -0.068 &  -0.007 & 0.028 & -0.010  & 0.034 \\ \hline
    Schnakenberg II & $D_2 =20, ~a=0.13,~b=1.4$ & $\sqrt{b_c-b}=0.241$  & -0.131 & -0.125 & -0.028 & -0.053 & 0.005 \\ \hline
  \end{tabular}
\end{center}
\caption{\label{tab2MSenv} A summary of marginal stability results (index $\MS$), numerical estimation (index $\RD$) and envelope equation results (index $\env$) of the key properties of the TW and the deposited pattern for $D_{1}=1$. Note that these models are defined in detail and explored in the main text and in the Appendix. We discuss each case via the relative error in both the front propagation speed and the wavenumber of the deposited pattern. In addition, we show the comparison of pattern amplitude and the role of the bifurcation parameter choice. 
}
\end{table}

 This study demonstrates the existence of a characteristic speed of pattern propagation determined from detailed reaction kinetics together with diffusion and thus is a hallmark of each Turing pattern formation system. Thus, in turn, each Turing system has an associated wave propagation speed that may be compared to experimental results 
in the same way as the length scale characteristic of the observed pattern \citep{miguez2006effect,konow2019turing,konow2021insights, glover2017hierarchical,glover2023developmental}. 
\bibliographystyle{cas-model2-names}

\bibliography{lit}

\begin{thebibliography}{28}
\expandafter\ifx\csname natexlab\endcsname\relax\def\natexlab#1{#1}\fi
\providecommand{\url}[1]{\texttt{#1}}
\providecommand{\href}[2]{#2}
\providecommand{\path}[1]{#1}
\providecommand{\DOIprefix}{doi:}
\providecommand{\ArXivprefix}{arXiv:}
\providecommand{\URLprefix}{URL: }
\providecommand{\Pubmedprefix}{pmid:}
\providecommand{\doi}[1]{\href{http://dx.doi.org/#1}{\path{#1}}}
\providecommand{\Pubmed}[1]{\href{pmid:#1}{\path{#1}}}
\providecommand{\bibinfo}[2]{#2}
\ifx\xfnm\relax \def\xfnm[#1]{\unskip,\space#1}\fi
\bibitem[{Avery et~al.(2023)Avery, Holzer and Scheel}]{avery2023pushed}
\bibinfo{author}{Avery, M.}, \bibinfo{author}{Holzer, M.},
  \bibinfo{author}{Scheel, A.}, \bibinfo{year}{2023}.
\newblock \bibinfo{title}{Pushed-to-pulled front transitions: continuation,
  speed scalings, and hidden monotonicty}.
\newblock \bibinfo{journal}{Journal of Nonlinear Science} \bibinfo{volume}{33},
  \bibinfo{pages}{102}.
\bibitem[{Ben-Jacob et~al.(1985)Ben-Jacob, Brand, Dee, Kramer and
  Langer}]{ben1985pattern}
\bibinfo{author}{Ben-Jacob, E.}, \bibinfo{author}{Brand, H.},
  \bibinfo{author}{Dee, G.}, \bibinfo{author}{Kramer, L.},
  \bibinfo{author}{Langer, J.}, \bibinfo{year}{1985}.
\newblock \bibinfo{title}{Pattern propagation in nonlinear dissipative
  systems}.
\newblock \bibinfo{journal}{Physica D: Nonlinear Phenomena}
  \bibinfo{volume}{14}, \bibinfo{pages}{348--364}.
\bibitem[{Berestycki and Hamel(2007)}]{berestycki2007generalized}
\bibinfo{author}{Berestycki, H.}, \bibinfo{author}{Hamel, F.},
  \bibinfo{year}{2007}.
\newblock \bibinfo{title}{Generalized travelling waves for reaction-diffusion
  equations}.
\newblock \bibinfo{journal}{Contemporary Mathematics} \bibinfo{volume}{446},
  \bibinfo{pages}{101--124}.
\bibitem[{Chomaz and Couairon(2000)}]{chomaz2000propagating}
\bibinfo{author}{Chomaz, J.M.}, \bibinfo{author}{Couairon, A.},
  \bibinfo{year}{2000}.
\newblock \bibinfo{title}{Propagating pattern selection and causality
  reconsidered}.
\newblock \bibinfo{journal}{Physical review letters} \bibinfo{volume}{84},
  \bibinfo{pages}{1910}.
\bibitem[{Dee and Langer(1983)}]{dee1983propagating}
\bibinfo{author}{Dee, G.}, \bibinfo{author}{Langer, J.}, \bibinfo{year}{1983}.
\newblock \bibinfo{title}{Propagating pattern selection}.
\newblock \bibinfo{journal}{Physical Review Letters} \bibinfo{volume}{50},
  \bibinfo{pages}{383}.
\bibitem[{Dee and van Saarloos(1988)}]{dee1988bistable}
\bibinfo{author}{Dee, G.}, \bibinfo{author}{van Saarloos, W.},
  \bibinfo{year}{1988}.
\newblock \bibinfo{title}{Bistable systems with propagating fronts leading to
  pattern formation}.
\newblock \bibinfo{journal}{Physical review letters} \bibinfo{volume}{60},
  \bibinfo{pages}{2641}.
\bibitem[{Glover et~al.(2023)Glover, Sudderick, Shih, Batho-Samblas, Charlton,
  Krause, Anderson, Riddell, Balic, Li et~al.}]{glover2023developmental}
\bibinfo{author}{Glover, J.D.}, \bibinfo{author}{Sudderick, Z.R.},
  \bibinfo{author}{Shih, B.B.J.}, \bibinfo{author}{Batho-Samblas, C.},
  \bibinfo{author}{Charlton, L.}, \bibinfo{author}{Krause, A.L.},
  \bibinfo{author}{Anderson, C.}, \bibinfo{author}{Riddell, J.},
  \bibinfo{author}{Balic, A.}, \bibinfo{author}{Li, J.}, et~al.,
  \bibinfo{year}{2023}.
\newblock \bibinfo{title}{The developmental basis of fingerprint pattern
  formation and variation}.
\newblock \bibinfo{journal}{Cell} \bibinfo{volume}{186},
  \bibinfo{pages}{940--956}.
\bibitem[{Glover et~al.(2017)Glover, Wells, Matth{\"a}us, Painter, Ho, Riddell,
  Johansson, Ford, Jahoda, Klika et~al.}]{glover2017hierarchical}
\bibinfo{author}{Glover, J.D.}, \bibinfo{author}{Wells, K.L.},
  \bibinfo{author}{Matth{\"a}us, F.}, \bibinfo{author}{Painter, K.J.},
  \bibinfo{author}{Ho, W.}, \bibinfo{author}{Riddell, J.},
  \bibinfo{author}{Johansson, J.A.}, \bibinfo{author}{Ford, M.J.},
  \bibinfo{author}{Jahoda, C.A.}, \bibinfo{author}{Klika, V.}, et~al.,
  \bibinfo{year}{2017}.
\newblock \bibinfo{title}{Hierarchical patterning modes orchestrate hair
  follicle morphogenesis}.
\newblock \bibinfo{journal}{PLoS biology} \bibinfo{volume}{15},
  \bibinfo{pages}{e2002117}.
\bibitem[{Hoyle(2006)}]{hoyle2006pattern}
\bibinfo{author}{Hoyle, R.}, \bibinfo{year}{2006}.
\newblock \bibinfo{title}{Pattern formation: an introduction to methods}.
\newblock \bibinfo{publisher}{Cambridge University Press}.
\bibitem[{Klika(2017)}]{klika2017significance}
\bibinfo{author}{Klika, V.}, \bibinfo{year}{2017}.
\newblock \bibinfo{title}{Significance of non-normality-induced patterns:
  Transient growth versus asymptotic stability}.
\newblock \bibinfo{journal}{Chaos: An Interdisciplinary Journal of Nonlinear
  Science} \bibinfo{volume}{27}.
\bibitem[{Kolmogorov(1937)}]{kolmogorov1937study}
\bibinfo{author}{Kolmogorov, A.N.}, \bibinfo{year}{1937}.
\newblock \bibinfo{title}{A study of the equation of diffusion with increase in
  the quantity of matter, and its application to a biological problem}.
\newblock \bibinfo{journal}{Moscow University Bulletin of Mathematics}
  \bibinfo{volume}{1}, \bibinfo{pages}{1--25}.
\bibitem[{Konow et~al.(2021)Konow, Dolnik and Epstein}]{konow2021insights}
\bibinfo{author}{Konow, C.}, \bibinfo{author}{Dolnik, M.},
  \bibinfo{author}{Epstein, I.}, \bibinfo{year}{2021}.
\newblock \bibinfo{title}{Insights from chemical systems into turing-type
  morphogenesis}.
\newblock \bibinfo{journal}{Philosophical Transactions of the Royal Society A}
  \bibinfo{volume}{379}, \bibinfo{pages}{20200269}.
\bibitem[{Konow et~al.(2019)Konow, Somberg, Chavez, Epstein and
  Dolnik}]{konow2019turing}
\bibinfo{author}{Konow, C.}, \bibinfo{author}{Somberg, N.H.},
  \bibinfo{author}{Chavez, J.}, \bibinfo{author}{Epstein, I.R.},
  \bibinfo{author}{Dolnik, M.}, \bibinfo{year}{2019}.
\newblock \bibinfo{title}{Turing patterns on radially growing domains:
  experiments and simulations}.
\newblock \bibinfo{journal}{Physical Chemistry Chemical Physics}
  \bibinfo{volume}{21}, \bibinfo{pages}{6718--6724}.
\bibitem[{Krause et~al.(2021)Krause, Gaffney, Maini and
  Klika}]{krause2021modern}
\bibinfo{author}{Krause, A.L.}, \bibinfo{author}{Gaffney, E.A.},
  \bibinfo{author}{Maini, P.K.}, \bibinfo{author}{Klika, V.},
  \bibinfo{year}{2021}.
\newblock \bibinfo{title}{Modern perspectives on near-equilibrium analysis of
  turing systems}.
\newblock \bibinfo{journal}{Philosophical Transactions of the Royal Society A}
  \bibinfo{volume}{379}, \bibinfo{pages}{20200268}.
\bibitem[{Lengyel and Epstein(1991)}]{lengyel1991modeling}
\bibinfo{author}{Lengyel, I.}, \bibinfo{author}{Epstein, I.R.},
  \bibinfo{year}{1991}.
\newblock \bibinfo{title}{Modeling of turing structures in the
  chlorite—iodide—malonic acid—starch reaction system}.
\newblock \bibinfo{journal}{Science} \bibinfo{volume}{251},
  \bibinfo{pages}{650--652}.
\bibitem[{Liu et~al.(2022)Liu, Maini and Baker}]{liu2022control}
\bibinfo{author}{Liu, Y.}, \bibinfo{author}{Maini, P.K.},
  \bibinfo{author}{Baker, R.E.}, \bibinfo{year}{2022}.
\newblock \bibinfo{title}{Control of diffusion-driven pattern formation behind
  a wave of competency}.
\newblock \bibinfo{journal}{Physica D: Nonlinear Phenomena}
  \bibinfo{volume}{438}, \bibinfo{pages}{133297}.
\bibitem[{M{\'\i}guez et~al.(2006)M{\'\i}guez, Dolnik, Munuzuri and
  Kramer}]{miguez2006effect}
\bibinfo{author}{M{\'\i}guez, D.G.}, \bibinfo{author}{Dolnik, M.},
  \bibinfo{author}{Munuzuri, A.P.}, \bibinfo{author}{Kramer, L.},
  \bibinfo{year}{2006}.
\newblock \bibinfo{title}{Effect of axial growth on turing pattern formation}.
\newblock \bibinfo{journal}{Physical review letters} \bibinfo{volume}{96},
  \bibinfo{pages}{048304}.
\bibitem[{Murray(2003)}]{murray2003mathematical}
\bibinfo{author}{Murray, J.D.}, \bibinfo{year}{2003}.
\newblock \bibinfo{title}{Mathematical Biology: II: Spatial Models and
  Biomedical Applications}. volume~\bibinfo{volume}{3}.
\newblock \bibinfo{publisher}{Springer}.
\bibitem[{Myerscough and Murray(1992)}]{myerscough1992analysis}
\bibinfo{author}{Myerscough, M.R.}, \bibinfo{author}{Murray, J.D.},
  \bibinfo{year}{1992}.
\newblock \bibinfo{title}{Analysis of propagating pattern in a chemotaxis
  system}.
\newblock \bibinfo{journal}{Bulletin of mathematical biology}
  \bibinfo{volume}{54}, \bibinfo{pages}{77--94}.
\bibitem[{Ponedel et~al.(2017)Ponedel, Kao and Knobloch}]{ponedel2017front}
\bibinfo{author}{Ponedel, B.C.}, \bibinfo{author}{Kao, H.C.},
  \bibinfo{author}{Knobloch, E.}, \bibinfo{year}{2017}.
\newblock \bibinfo{title}{Front propagation in weakly subcritical
  pattern-forming systems}.
\newblock \bibinfo{journal}{Physical Review E} \bibinfo{volume}{96},
  \bibinfo{pages}{032208}.
\bibitem[{Rovinsky and Menzinger(1992)}]{rovinsky1992chemical}
\bibinfo{author}{Rovinsky, A.B.}, \bibinfo{author}{Menzinger, M.},
  \bibinfo{year}{1992}.
\newblock \bibinfo{title}{Chemical instability induced by a differential flow}.
\newblock \bibinfo{journal}{Physical Review Letters} \bibinfo{volume}{69},
  \bibinfo{pages}{1193}.
\bibitem[{Sandstede and Scheel(2000)}]{sandstede2000absolute}
\bibinfo{author}{Sandstede, B.}, \bibinfo{author}{Scheel, A.},
  \bibinfo{year}{2000}.
\newblock \bibinfo{title}{Absolute and convective instabilities of waves on
  unbounded and large bounded domains}.
\newblock \bibinfo{journal}{Physica D: Nonlinear Phenomena}
  \bibinfo{volume}{145}, \bibinfo{pages}{233--277}.
\bibitem[{Schnakenberg(1979)}]{schnakenberg1979simple}
\bibinfo{author}{Schnakenberg, J.}, \bibinfo{year}{1979}.
\newblock \bibinfo{title}{Simple chemical reaction systems with limit cycle
  behaviour}.
\newblock \bibinfo{journal}{Journal of Theoretical Biology}
  \bibinfo{volume}{81}, \bibinfo{pages}{389--400}.
\bibitem[{Sherratt et~al.(2014)Sherratt, Dagbovie and
  Hilker}]{sherratt2014mathematical}
\bibinfo{author}{Sherratt, J.A.}, \bibinfo{author}{Dagbovie, A.S.},
  \bibinfo{author}{Hilker, F.M.}, \bibinfo{year}{2014}.
\newblock \bibinfo{title}{A mathematical biologist’s guide to absolute and
  convective instability}.
\newblock \bibinfo{journal}{Bulletin of mathematical biology}
  \bibinfo{volume}{76}, \bibinfo{pages}{1--26}.
\bibitem[{Tarumi and Mueller(1989)}]{tarumi1989wavelength}
\bibinfo{author}{Tarumi, K.}, \bibinfo{author}{Mueller, E.},
  \bibinfo{year}{1989}.
\newblock \bibinfo{title}{Wavelength selection mechanism in the
  gierer-meinhardt model}.
\newblock \bibinfo{journal}{Bulletin of mathematical biology}
  \bibinfo{volume}{51}, \bibinfo{pages}{207--216}.
\bibitem[{Tobias et~al.(1998)Tobias, Proctor and
  Knobloch}]{tobias1998convective}
\bibinfo{author}{Tobias, S.M.}, \bibinfo{author}{Proctor, M.R.E.},
  \bibinfo{author}{Knobloch, E.}, \bibinfo{year}{1998}.
\newblock \bibinfo{title}{Convective and absolute instabilities of fluid flows
  in finite geometry}.
\newblock \bibinfo{journal}{Physica D: Nonlinear Phenomena}
  \bibinfo{volume}{113}, \bibinfo{pages}{43--72}.
\bibitem[{Van~Saarloos(1988)}]{vansaarloos1988front}
\bibinfo{author}{Van~Saarloos, W.}, \bibinfo{year}{1988}.
\newblock \bibinfo{title}{Front propagation into unstable states: marginal
  stability as a dynamical mechanism for velocity selection}.
\newblock \bibinfo{journal}{Physical Review A} \bibinfo{volume}{37},
  \bibinfo{pages}{211}.
\bibitem[{Van~Saarloos(2003)}]{vansaarlos2003front}
\bibinfo{author}{Van~Saarloos, W.}, \bibinfo{year}{2003}.
\newblock \bibinfo{title}{Front propagation into unstable states}.
\newblock \bibinfo{journal}{Physics Reports} \bibinfo{volume}{386},
  \bibinfo{pages}{29--222}.

\end{thebibliography}



\newpage
\appendix

\section{CDIMA reaction kinetics}\label{App.OtherCases}

In the main text, we provide a detailed illustration of the analytical results for both the marginal stability and envelope equation method with Schnakenberg reaction kinetics with particular choices of parameter values. For instance, for the case labelled as Schnakenberg I, we have the parameter choice $D_{1}=1$, $D_{2}=20$, $a=0.05$, $b=1.4$ and with the domain size $L=200$, the evolution of the localised initial condition with $\rho=3$ results in the formation of a pattern behind a front travelling with an approximately constant speed, see Fig \ref{fig1:Sch-PDEsolu}, which can be associated with the Turing space and the bifurcation point given in Fig.~\ref{fig2:Sch-Tspace}. In this Supplementary section, we present analogous results for the CDIMA reaction kinetics not presented in the main text, except via the summary statistics of Table \ref{tab1MS} and \ref{tab2MSenv}.

\subsection{Preliminaries}
We consider the reaction-diffusion problem \eqref{eq:RD0} with the, albeit simplified, modelling representation of the CDIMA (chlorine dioxide-iodine-malonic acid) reaction kinetics taken from \citet{konow2019turing}, which is based on the two-variable version of the kinetics \citep{lengyel1991modeling}, that is  
 CDIMA reaction kinetics \eqref{eq.CDIMA} on a one-dimensional spatial domain with zero-flux boundary conditions.

\begin{figure}
\begin{subfigure}{.48\textwidth}
  \centering
    \includegraphics[width=.95\linewidth]{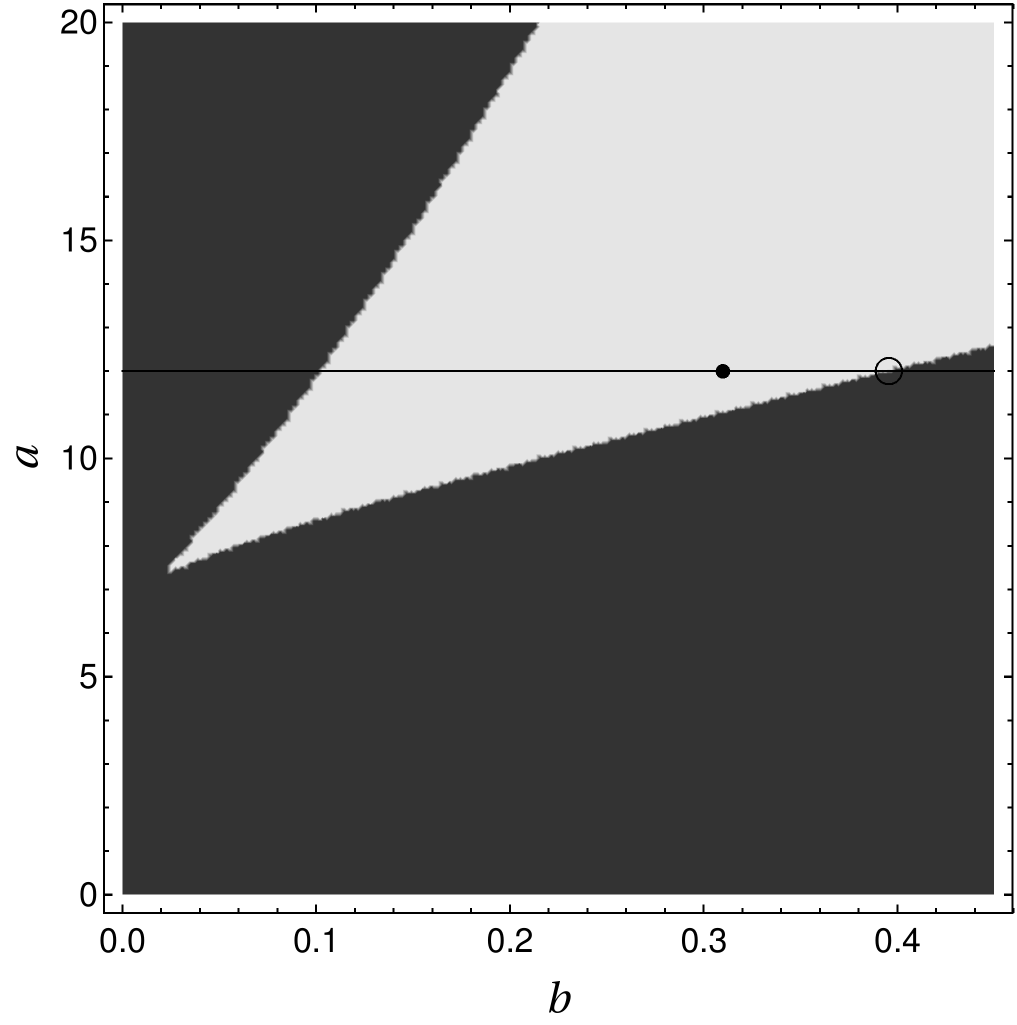}
    \caption{The Turing space, the set of points which allow Turing instability, for CDIMA kinetics in $a,~b$ parameters, is highlighted in grey, given the other parameter values are $D_1 =1,~ D_2 =\mu$, $\mu=50$ and a 1D domain with size $L=200$. The full dot represents the chosen point in the $a,~b$ parameter space for numerical solution of the full problem (CDIMA I). The open circle denotes the closest bifurcation point $b_c$ (corresponding to $\alpha=0$) for the chosen bifurcation parameter $\alpha=b_c-b$. }
    \label{figA1:CDIMA-Tspace}
  \end{subfigure}
    \hspace{0.3cm}
  \begin{subfigure}{.48\textwidth}
   \centering
\includegraphics[width=\linewidth]{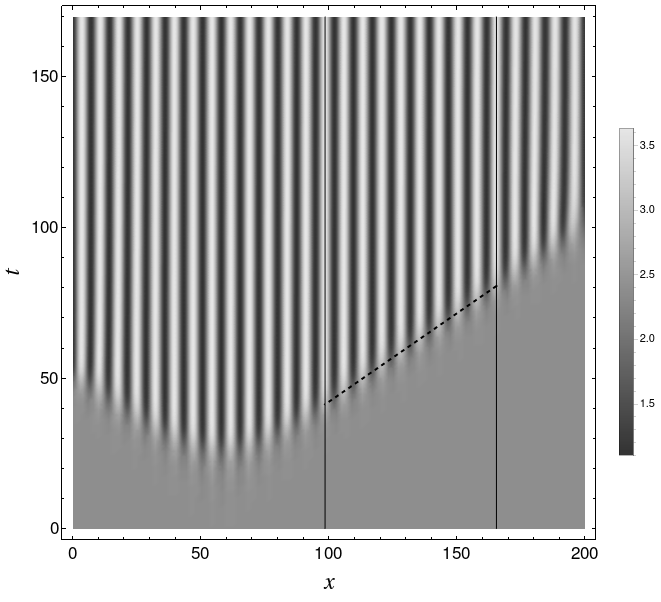}
   \caption{We initiate the simulation by a local perturbation in the second component as shown in the left panel. The density plot of the first component of the solution, $u_1$, shows the resulting transient wave of patterning spreading across the domain in both directions (left and right) with an approximately constant velocity, prior to the establishment of a stationary Turing pattern. After an initial transient period, we can indeed observe the asymptotic establishment of a constant travelling wave speed as indicated by the straight dashed line.}
     \label{figA1-CDIMA_illu}
   \end{subfigure}
    \caption{\label{fig:A1}
 Illustration of pattern formation behind a wave travelling at a constant speed after a localised perturbation to the steady state. We consider the CDIMA system, Eqn. \eqref{eq:RD0}, with zero-flux boundary conditions and parameter values $D_1 =1,~ D_2 =\mu$, $a=12,~b=0.31,~\mu=50$ and a 1D domain with size $L=200$ (CDIMA I), using an initial perturbation of the form of that presented in Fig ~\ref{fig1:IC} with $\rho=3$. The profile of $u_{2}$ is in phase with the profile of $u_{1}$.}
\end{figure}

We take the parameters as $D_1 =1,~ D_2 =\mu$, $a=12,~b=0.31, ~\mu=50$ and a 1D domain with size $L=200$ (CDIMA I).  This set of parameters is within the Turing space as indicated in Figure \ref{figA1:CDIMA-Tspace}, where we plot the Turing space in the $a,~b$ parameter space for $D_1$, $D_2$ fixed at $1, 50$ respectively. We also highlight the localisation of the nearest bifurcation point $b_c=0.396$ for the chosen bifurcation parameter $\alpha=b_c-b$, which leads to $\epsilon=\sqrt{\alpha}=0.293$. 
With this choice of parameter values, the evolution of the localised initial condition with $\rho=3$ results in the formation of a pattern behind a front travelling with an approximately constant speed, see Fig. \ref{figA1-CDIMA_illu}.  In addition, from the numerical solution determined in this manner, we may extract all the observed characteristics from the (numerical) localisation of the highlighted maxima of the pattern, as in Figure \ref{figA1-CDIMA_illu} with:
i) the amplitude $A_\RD=1.223$; ii) the pattern   wavenumber $k_\RD=0.835$, denoting $2\pi$ divided by the distance between the neighbouring maxima; and iii) the speed of the travelling wave $c_\RD=1.626$ from numerically calculating  the times when the travelling pattern has reached half of the final amplitude at the two highlighted locations.

\subsection{Envelope method}\label{App.OtherCases_Envelope}

For the CDIMA kinetics, similar details are given here, analogously to the above-mentioned example in the main text: Schnakenberg kinetics with the choice of Schnakenberg I parameter values.

\begin{figure}
\centering
\begin{subfigure}{.48\textwidth}
    \centering
    \includegraphics[width=.95\linewidth]{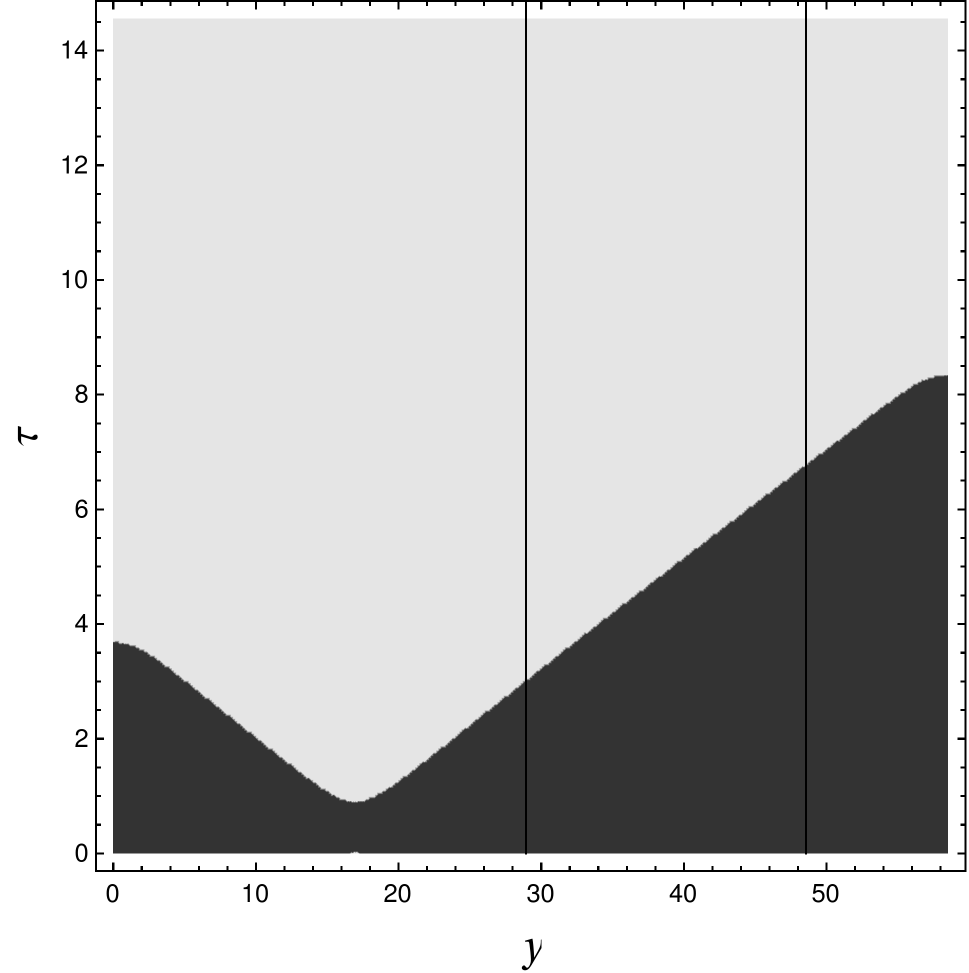}
    \caption{Density plot of the numerical solution to the envelope equation Eq. \eqref{eq.CDIMAenvExample}. The black region is a set of points $(y,\tau)$ where  $A(\tau,y)<A_\env/2$, while the white region shows its complement.}
    \label{fig:CDIMA-densityEnvelope_halfA}
  \end{subfigure}
  \hspace{0.3cm}
\begin{subfigure}{.48\textwidth}
    \centering
    \includegraphics[width=.95\linewidth]{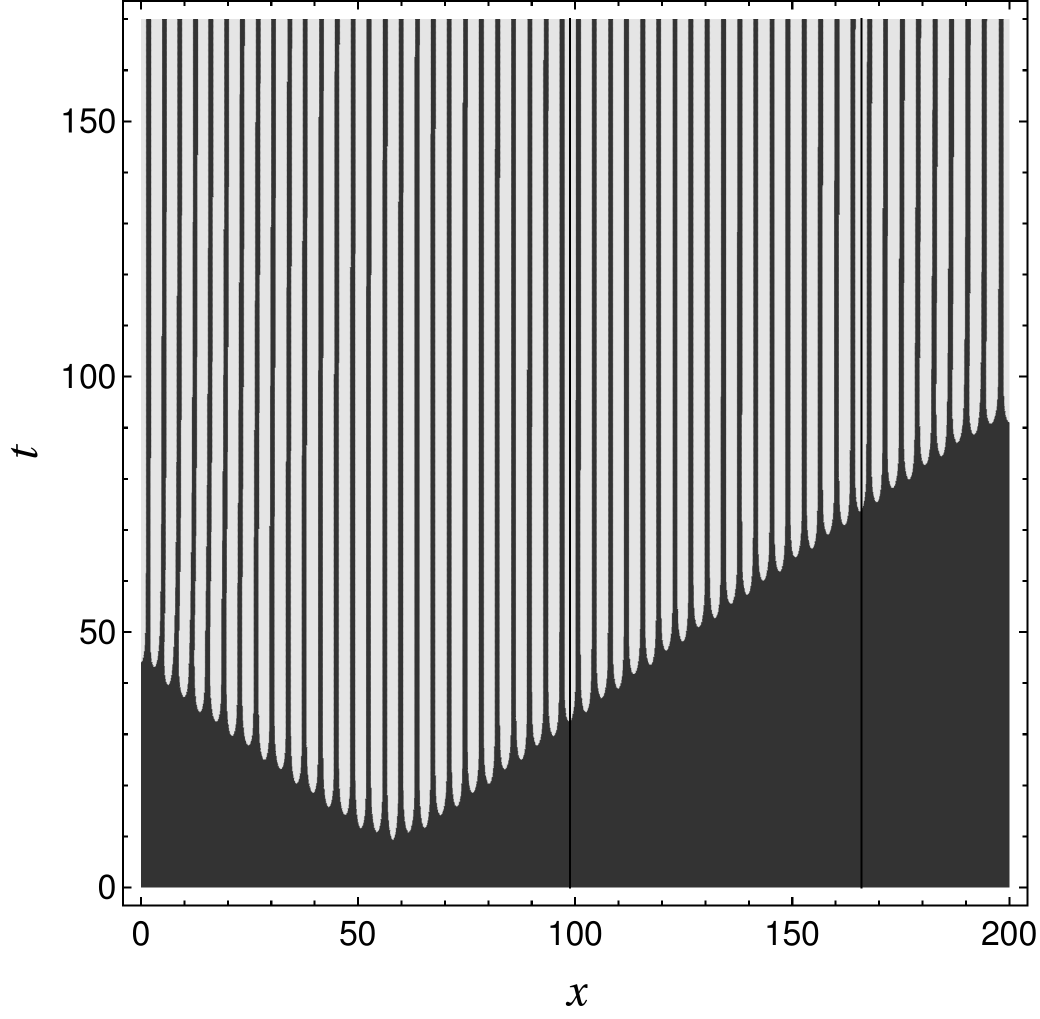}
    \caption{Density plot of $u_1$ of the numerical solution to the full problem.  The black region is a set of points $(x,t)$ where  $A(\tau,y)<A_\RD/2$ while the white region shows its complement.}
    \label{fig:CDIMA-densityRD_halfA}
\end{subfigure}
\caption{CDIMA kinetics  with $D_1 =1,~ D_2 =\mu$, $a=12,~b=0.31,~\mu=50$ and a 1D domain with size $L=200$ (CDIMA I) using the initial conditions given by Fig \ref{fig1:IC} with $\rho=3$. The numerical solution for the derived envelope equation is presented in panel (a), and for the full problem, panel (b). Again the vertical black lines highlight the positions of cross-sections for determining travelling wave speed and amplitudes. Note the similarity in both the time and location of the pattern initiation and also the development of the spatial pattern behind a front travelling with a fixed velocity (neglecting the initial transient behaviour and boundary effects). We plot the solution to the envelope equation in the corresponding coordinates $\tau=\epsilon^{2}t$, $y=\epsilon x$ (with $\epsilon=0.293$ for the listed parameter values). Hence the two plots are directly comparable and the two solutions are almost overlapping.}
\label{fig2A}
\end{figure}

\paragraph{CDIMA I.}
We consider CDIMA kinetics with $D_1 =1,~ D_2 =\mu$, $a=12,~b=0.31,~\mu=50$ and a 1D domain with size $L=200$ (CDIMA I). Using the above analysis, we obtain, for this choice of reaction kinetics and parameter values, the following form of the envelope equation
\begin{equation} \label{eq.CDIMAenvExample}
  \frac{\partial A}{\partial \tau}=3.496\frac{\partial^2 A}{\partial y^2} + 2.161 A-0.212 A^3,
\end{equation}
and that $\epsilon=0.293$. Hence, we may simply read out the predicted amplitude of the pattern $A_\env=2\epsilon A^*=1.869$ and, using Eq. \eqref{eq:NatSpeedOfPattern}, we have an estimate for the travelling wave speed $c_\env=1.609$. The wavenumber $k_\env$ is equal to the critical wavenumber $k_c$, Eq. \eqref{eq:kc}, and has the value $k_\env=k_c=0.915$. In addition, we solve this scalar reaction diffusion equation with the same zero flux boundary conditions and the same initial condition as above in the full problem, that is, a localised perturbation of the same magnitude, $\rho=3$, and location $x=60$ about the unstable trivial homogeneous steady state. 
Plotting this solution reveals a good qualitative agreement with the solution to the full problem, see Figures \ref{fig:CDIMA-densityEnvelope_halfA}, \ref{fig:CDIMA-densityRD_halfA}, when we plot the solutions in their natural coordinates.

In addition, from the numerical solution to the envelope equation we determine the numerically observed speed of the travelling wave $c_\env^\num$ as described above in the two highlighted locations giving $c_\env^\num=1.562$, showing a good match with the critical wave speed determined analytically from Kolmogorov asymptotic arguments.

\begin{figure}
\centering
\begin{subfigure}{.47\textwidth}
    \centering
    \includegraphics[width=.95\linewidth]{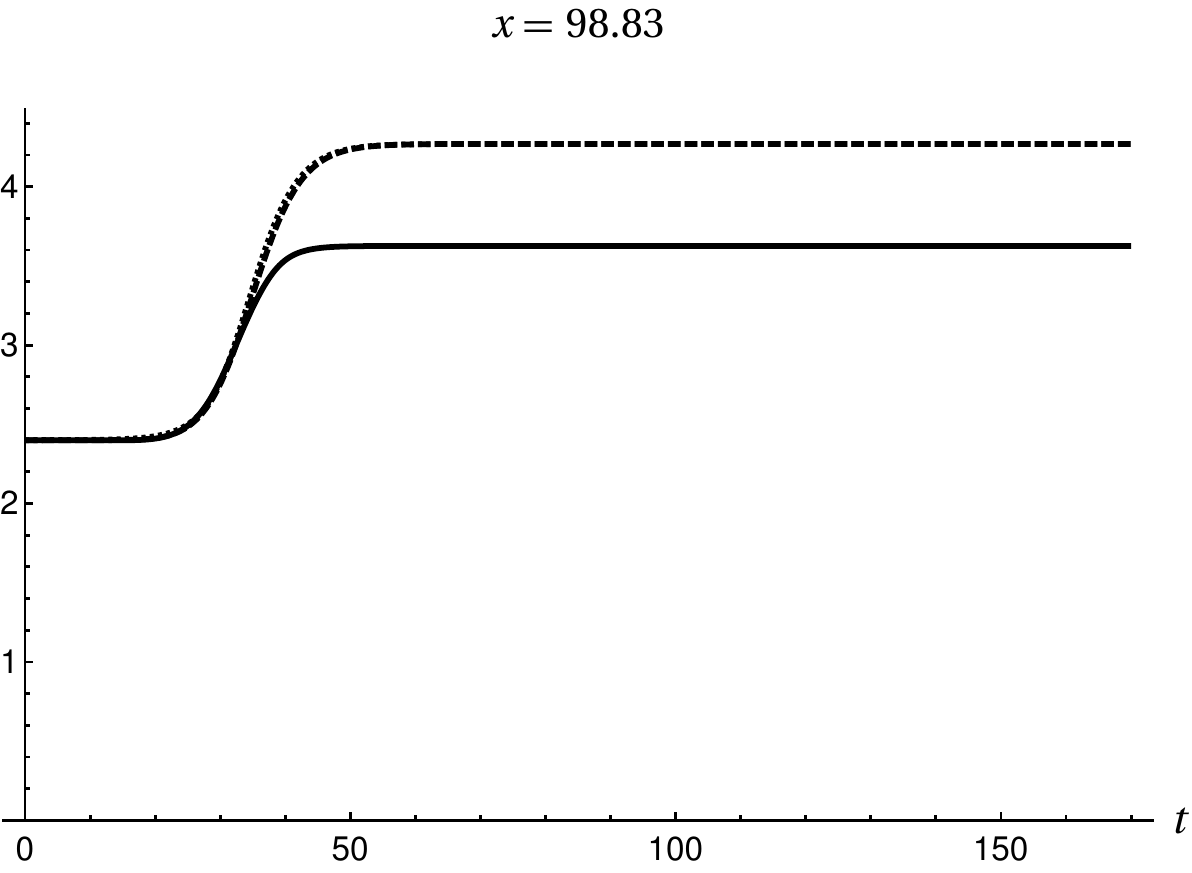}
    \caption{The temporal profile for the numerical solutions and the estimated analytical profile at the first location given by $x=98.83$ in Figs. \ref{figA1-CDIMA_illu} and \ref{fig2A}b. All solutions are shown in the original $t,~x$ variables.}
    \label{fig:CDIMA-EnvVsPDE-early}
\end{subfigure}
\hspace{0.5cm}
\begin{subfigure}{.47\textwidth}
    \includegraphics[width=.95\linewidth]{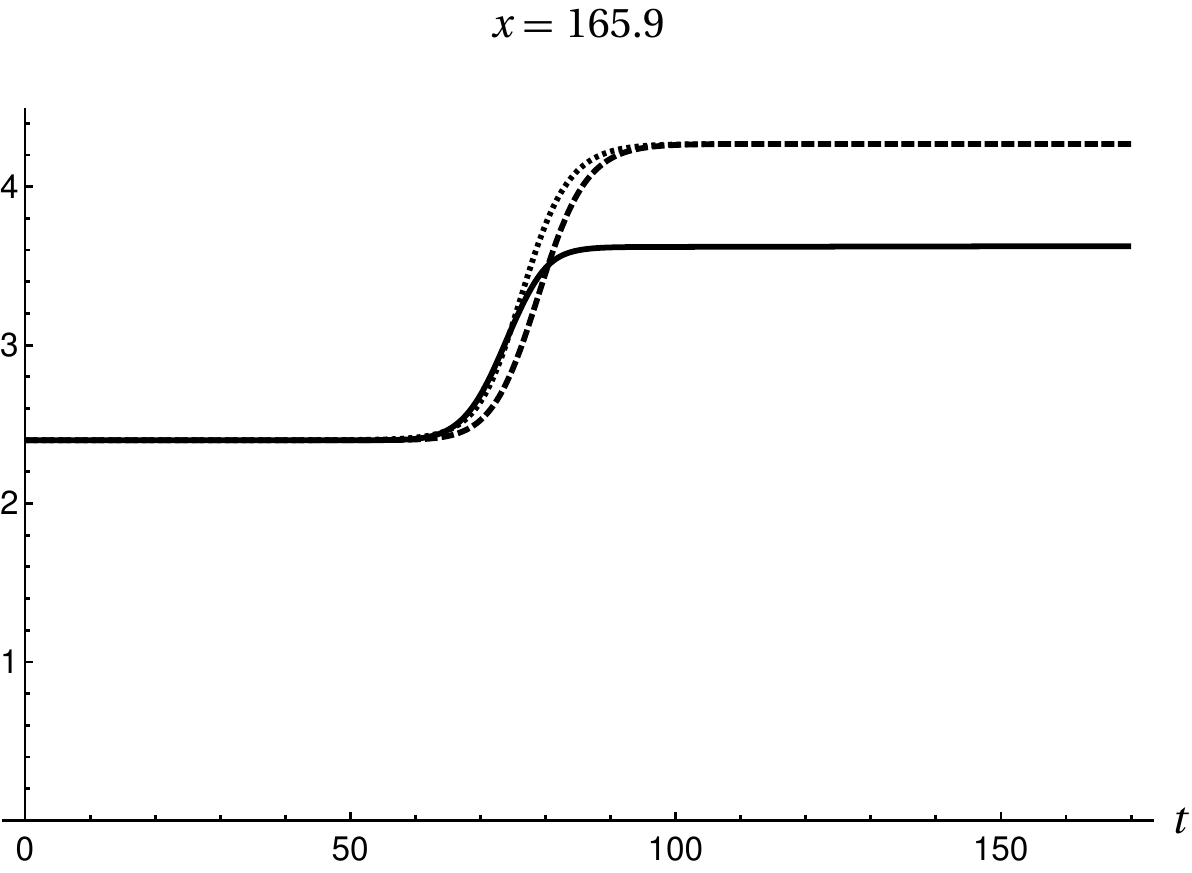}
    \caption{The temporal profile for the numerical solutions and the estimated analytical profile at the second location given by $x=165.9$ in Figs. \ref{figA1-CDIMA_illu} and \ref{fig2A}b. All solutions are shown in the original $t,~x$ variables.}
    \label{fig:CDIMA-EnvVsPDE-late}
\end{subfigure}
\caption{CDIMA kinetics with $D_1 =1,~ D_2 =\mu$, $a=12,~b=0.31,~\mu=50$ and a 1D domain with size $L=200$ (CDIMA I) using the initial conditions given by Fig \ref{fig1:IC} with $\rho=3$. Temporal profiles of the solutions at the highlighted positions - vertical black lines in the preceding density plots, namely in Figs. \ref{figA1-CDIMA_illu} and \ref{fig2A}b. The full curve corresponds to the solution $u_1(t,x)$ to the full problem. The dashed curve is the numerical solution to the corresponding envelope problem $(u^*)_1+2\epsilon A(\epsilon^2 t,\epsilon x)$, and the dotted curve (significantly overlapping with the dashed) is the shifted in time estimated analytical profile $(u^*)_1+2\epsilon A\left(\frac{\epsilon}{\sqrt{d}} (x-c_\env \epsilon (t+29))\right)$ from Eq. \eqref{eq:TWprofileApprox}. This shift in the plot of the analytical profile is necessary because it was obtained from the phase space in TW coordinates and therefore is subject to the translational invariance of the travelling wave which therefore must be fixed, see text for more details. As one can observe from the comparison of the two panels, all the three speeds $c_\RD,~c_\env,~c_\env^\num$ of the travelling wave are similar and the approximate analytical profile matches the numerically calculated ones, see the text for more details. Finally, note that there is a disparity in the predicted amplitude of the pattern, the error being $53\%$ of the numerically calculated amplitude from the full problem. We expect this error to reduce with $\epsilon$.}
\label{fig4}
\end{figure}

Finally, we plot the solutions, see Figures \ref{fig:CDIMA-EnvVsPDE-early}, \ref{fig:CDIMA-EnvVsPDE-late}, to both the full problem and the envelope equation for the two highlighted cross-sections in Fig. \ref{figA1-CDIMA_illu} corresponding to $x=98.83$ and $x=165.9$. In this way we can compare the predicted travelling wave profile together with its amplitude and velocity. We also plot the asymptotic estimate of the travelling wave profile, \eqref{eq:TWprofileApprox}, shifted in time by $t_0$ to match the arrival at the first cross-section with the numerical solution to the envelope equation. 
This time shift $t_0$ is necessary as the determination of the travelling wave profile does not consider initiation and development of the front and thus is free up to a translational shift, which we fix by specifying $t_0$. Its value is determined manually by choosing a value which results in the best match (visually overlapping) between the analytical profile, eq \eqref{eq:TWprofileApprox} and plotted as a dotted curve, together with a plot of the numerical solution for the envelope problem at the first cross-section (dashed). In particular, we plot the function $(u^*)_1+2\epsilon A(\xi)$ where $A(\xi)$ is the identified asymptotic approximation for the amplitude in \eqref{eq:TWprofileApprox} evaluated at the point $\xi(t,x)=\frac{\epsilon}{\sqrt{d}} (x-c_\env \epsilon (t+t_0))$. 
Hence, there is a single fitting parameter,  the time shift $t_0$. Therefore, the precision of the analytical estimate of the front velocity $c_\env$ is visually immediate as the difference between the times of arrival of the waves (full, dashed and dotted curves) at the second cross-section profile, Fig \ref{fig:CDIMA-EnvVsPDE-late}.

We also show a solution to the full RD problem where there is no need to provide a time shift correction. Thus, from the figure we are able to determine the velocities observed in all three approaches: (i) the full reaction diffusion system with numerically estimate speed, $c_\RD$; (ii) the  analytical estimate for the asymptotic travelling wave speed from the envelope equation $c_\env$;  and (iii) the numerically estimated asymptotic travelling wave speed from the numerical solution of the envelope equation,   $c_\env^{\mathrm{num}}$. In addition, we can also assess   the closeness of fit for the front profiles and the time taken for the pattern to develop from the small localised disturbance.

\paragraph{CDIMA II.}
We consider CDIMA kinetics with parameter values $D_1=1,~D_2=2\mu,~a=10.5,~b=0.4,~\mu=13$ and $L=200$ referred to as CDIMA II. 

In this situation, from the envelope equation analysis, we obtain
\begin{equation*}
  \frac{\partial A}{\partial \tau}=3.644\frac{\partial^2 A}{\partial y^2} + 1.421 A-0.582 A^3.
\end{equation*}
and $\epsilon=0.346$, $k_\env=k_c=0.843$, $A_\env=1.082$, $c_\env=1.575$, $c_\env^\num=1.477$.

\begin{figure}
\centering
\begin{subfigure}{.48\textwidth}
    \centering
    \includegraphics[width=.95\linewidth]{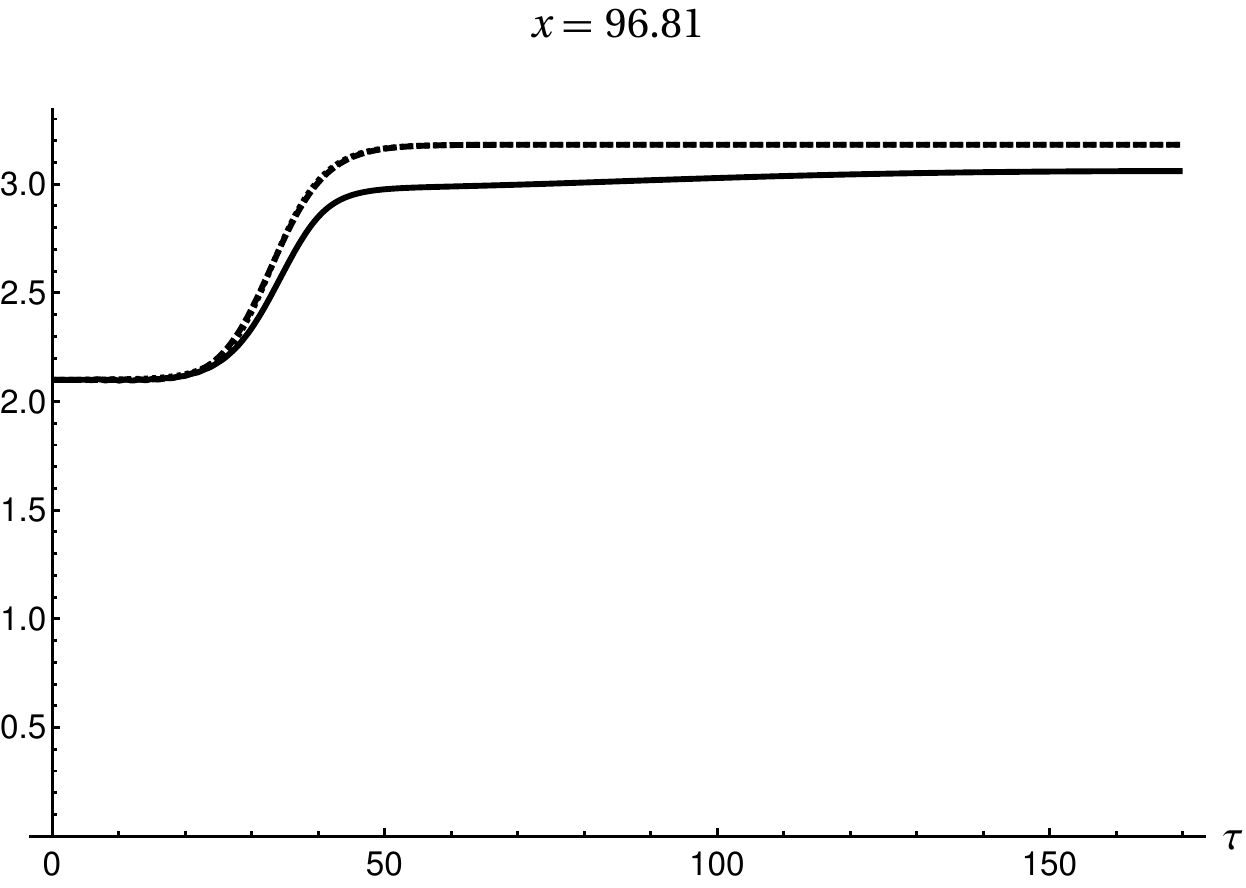}
    \caption{The temporal profile for the numerical solutions and the estimated analytical profile at the first location given by $x=96.81$. All solutions are shown in the original $t,~x$ variables.}
    \label{fig:CDIMADas-EnvVsPDE-early}
  \end{subfigure}
  \hspace{0.3cm}
\begin{subfigure}{.48\textwidth}
    \centering
    \includegraphics[width=.95\linewidth]{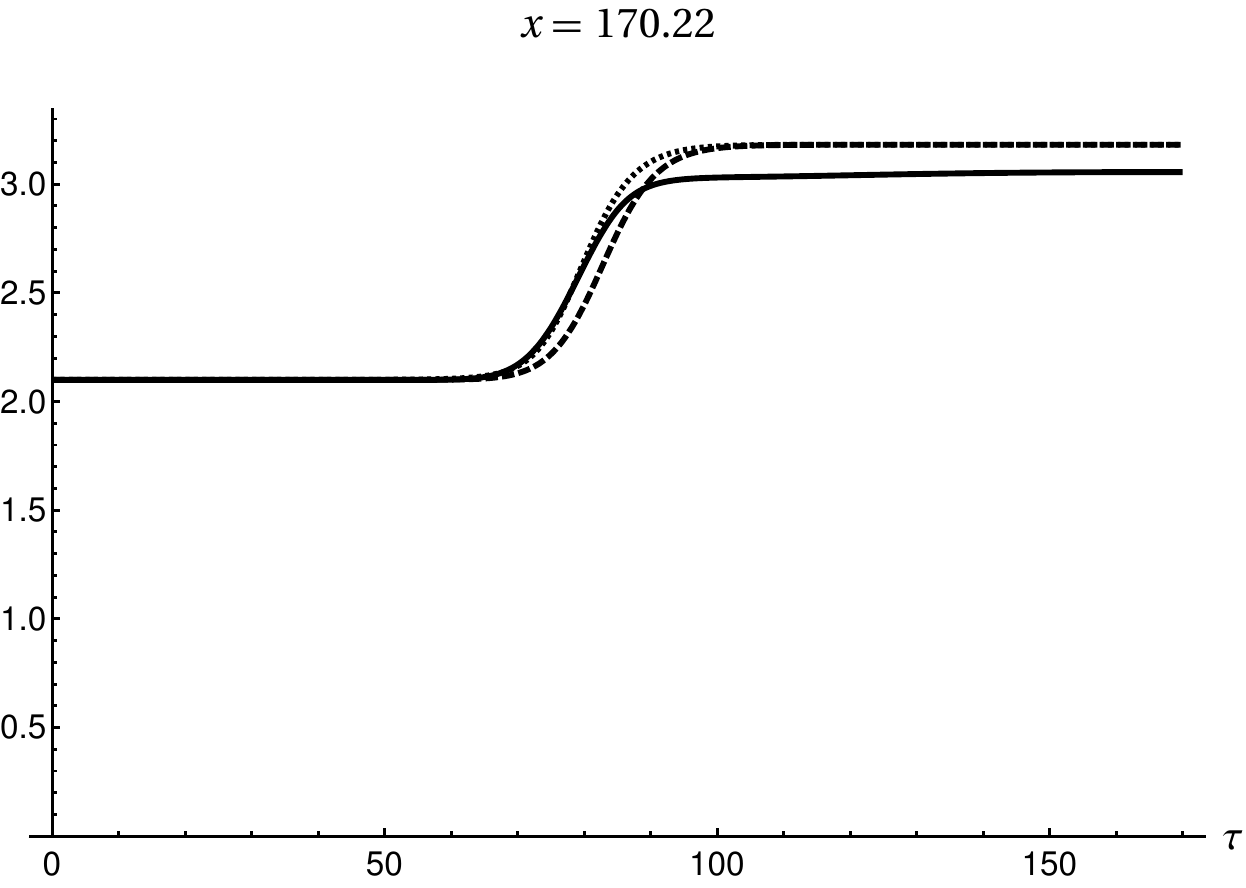}
    \caption{The temporal profile for the numerical solutions and the estimated analytical profile at the second location given by $x=170.22$. All solutions are shown in the original $t,~x$ variables.}
    \label{fig:CDIMAII-EnvVsPDE-late}
\end{subfigure}
\caption{CDIMA kinetics with $D_1 =1,~ D_2 =2\mu$, $a=10.5,~b=0.4,~\mu=13$ and a 1D domain with size $L=200$ (CDIMA II) using the initial conditions given by Fig \ref{fig1:IC} with $\rho=3$. Temporal profiles for the solutions at the highlighted positions. 
  The full curve corresponds to the solution $u_1(t,x)$ to the full problem. The dashed curve is the numerical solution to the corresponding envelope problem $(u^*)_1+2\epsilon A(\epsilon^2 t,\epsilon x)$, and the dotted curve (significantly overlapping with the dashed curve) is the shifted in time estimated analytical profile $(u^*)_1+2\epsilon A\left(\frac{\epsilon}{\sqrt{d}} (x-c_\env \epsilon (t+31))\right)$ from Eq. \eqref{eq:TWprofileApprox}. This shift in the plot of the analytical profile is necessary because it was obtained from the phase space in TW coordinates and therefore does not reflect the dynamics of travelling wave formation, see text for more details. Note that all the three speeds $c_\RD,~c_\env,~c_\env^\num$ of the travelling wave are similar and the approximate analytical profile matches the numerically calculated ones, see text for more details. Finally, note that there is a disparity in the predicted amplitude of the pattern but it is smaller than in the main text, the error being $13\%$.}
\label{fig:CDIMADas-EnvVsPDE}
\end{figure}

In Figure \ref{fig:CDIMADas-EnvVsPDE} we again compare the travelling wave nature, profile, amplitude, and speed as a solution to the full problem and envelope equation, while we also plot the estimated analytical profile of the travelling wave. We again determine the shift $t_0$ of the analytical amplitude profile as described above in CDIMA I. As a result, we can directly compare the travelling wave velocity and the profile from the three considered approaches, see Figure \ref{fig:CDIMADas-EnvVsPDE} and the discussion in the main text.

\paragraph{CDIMA III.}
As a final CDIMA example, we consider parameter values $D_1=1,~D_2=\mu,~a=12,~b=0.38,~\mu=50$ and $L=200$ (CDIMA III). This is a variation of CDIMA I where we have moved the bifurcation parameter $b$ closer to the bifurcation point $b_c$ while keeping the remaining parameters fixed.

The envelope equation analysis yields
\begin{equation*}
  \frac{\partial A}{\partial \tau}=3.496\frac{\partial^2 A}{\partial y^2} + 2.161 A-0.213 A^3.
\end{equation*}
and $\epsilon=0.125$, $k_\env=0.915$, $A_\env=0.796$, $c_\env=0.686$, $c_\env^\num=0.645$.
We do not plot the solution to this problem but we list the key characteristics in Table \ref{tab2MSenv}.

\subsection{Other choices of bifurcation parameters} \label{App.OtherAlphasEnv}

As mentioned in the main text, the choice of the bifurcation parameter may play a role in the weakly nonlinear analysis of behaviour near the bifurcation point.

We remark that the choice $\alpha=(a-a_c)/a_c$ in CDIMA I model, where $a_c=11.058$ is the critical parameter value, gives the following results
\begin{equation*}
  \frac{\partial A}{\partial \tau}=4.494\frac{\partial^2 A}{\partial y^2} + 2.578 A-0.209 A^3.
\end{equation*}
and $\epsilon=0.292$, $A_\env=2.052$, $c_\env=1.987$, $c_\env^\num=1.885$. The wavenumber is affected by this change in $\alpha$ as it follows from the properties of ${\bf J}$ at the bifurcation point $a=a_c$. In particular, with the considered choice of the bifurcation parameter $\alpha$, the wavenumber is approximated as $k_c=0.924$.

Similarly, the choice $\alpha=(\mu_c-\mu)/\mu_c$ in CDIMA I model, where $\mu_c=63.813$ is the bifurcation point, yields
\begin{equation*}
  \frac{\partial A}{\partial \tau}=3.496\frac{\partial^2 A}{\partial y^2} + 0.855 A-0.212 A^3.
\end{equation*}
and $\epsilon=0.465$, $A_\env=1.869$, $c_\env=1.609$, $c_\env^\num=1.521$. In this situation, the wavenumber is estimated as $k_c=0.915$.

We do not plot the solutions with these choices but we list the key characteristics for comparison in the first three rows of Table \ref{tab2MSenv}. One can see that even though the estimates are carried out for exactly the same system (the same kinetics, diffusion and parameter values), the choice of the bifurcation parameter $\alpha$ affects the accuracy of predictions from the envelope approach.




\subsection{Comparison of all methods}

\begin{figure}
\centering
\begin{subfigure}{.48\textwidth}
    \centering
    \includegraphics[width=.95\linewidth]{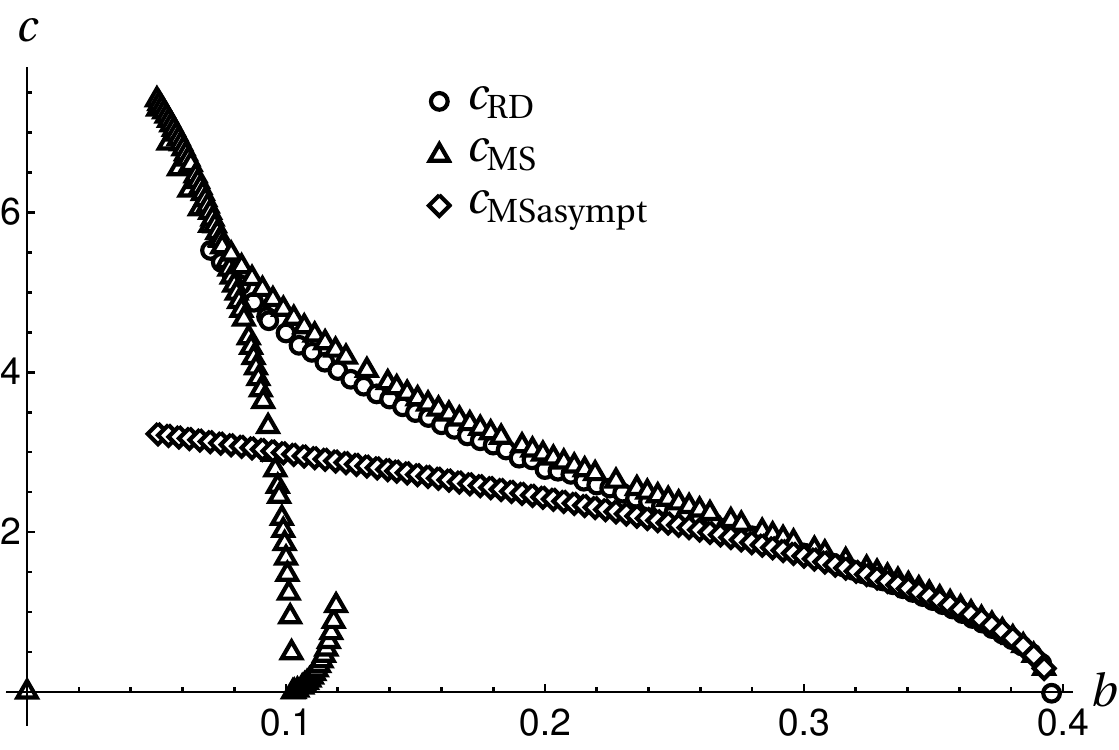}
    \caption{Convergence of all the methods to the same square root behaviour close to the bifurcation point $b_c=0.396$ is apparent. Note that the marginal stability approach shows multiple roots (multiple values of predicted TW speeds) as there are two branches of $c_\MS$ for $b<0.12$.}
    \label{fig:BifDiagCDIMA}
  \end{subfigure}
  \hspace{0.3cm}
\begin{subfigure}{.48\textwidth}
    \centering
    \includegraphics[width=.95\linewidth]{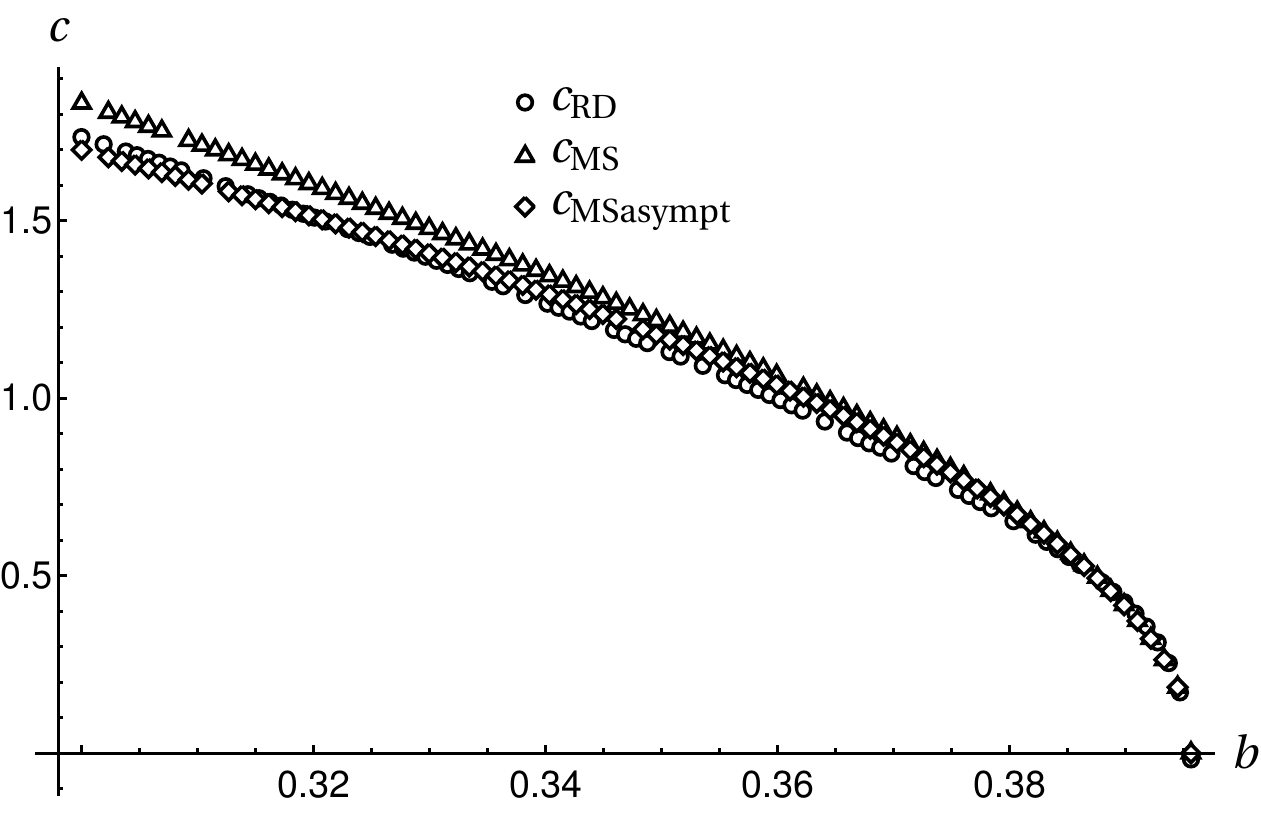}
    \caption{A magnification of the figure in the left panel closer to the bifurcation point showing the convergence of the marginal stability asymptotics to the marginal stability criterion results.}
    \label{fig:BifDiagCDIMAzoom}
\end{subfigure}
\caption{Travelling wave speed (given in $x,~t$ units)) as predicted from (i) the full solution of RD problem with CDIMA kinetics, $c_\RD$; (ii) from the marginal stability conditions, $c_\MS$; and (iii) from the marginal stability asymptotics, $c_{\rm MSasympt}$. Parameter values were chosen as $D_1 =1,~ D_2 =\mu$, $a=12$, $\mu=50$ and 1D domain with size $L=200$ (CDIMA I). Note that the choice of the magnitude of the localised initial condition in the finite domains is important as demonstrated in Fig \ref{fig3:Sch-CvsIC} and we consider $\rho=3$. Note that the marginal stability approach can yield multiple solutions, as illustrated here for $b\in(0.08,0.12)$ where two branches appear.} 
\label{fig:CDIMA-Cvsb}
\end{figure}

As a final illustration, we show the comparison of marginal stability criterion results (and its multiple roots), envelope method, and the characteristics stemming from the numerical solution to the full problem, as an analogue to Fig \ref{fig4:Cvsb} for the Schnakenberg case in the main text. In Fig. \ref{fig:CDIMA-Cvsb}, we plot the TW  velocity in $x,~t$ variables. We again observe a behaviour corresponding to the square root of the distance from the bifurcation point in its neighbourhood and that there is a match between the asymptotic expression for the travelling wave speed close to the bifurcation point and the numerical solution of the marginal stability criterion. 
Finally, note once more that the equations for marginal stability may have multiple solutions, where multiple branches of $c_\MS$ appear (even close to the bifurcation point). 

\end{document}